\documentclass[conference]{IEEEtran}
\usepackage{cite}
\usepackage{amsmath,amssymb,amsfonts}
\usepackage{algorithmic}
\usepackage{graphicx}
\usepackage{textcomp}
\usepackage{xcolor}
\usepackage{float}
\usepackage{hyperref}
\hypersetup{colorlinks=true, 
breaklinks=true, 
urlcolor= black, 
linkcolor= black,
citecolor= black 
} 

\usepackage{cleveref} 
\crefname{figure}{fig.}{fig.}
\Crefname{figure}{Fig.}{Fig.}
\crefname{equation}{}{}
\Crefname{equation}{}{}

\usepackage{braket}

\usepackage{scalerel,stackengine}
\stackMath
\newcommand\reallywidehat[1]{%
\savestack{\tmpbox}{\stretchto{%
  \scaleto{%
    \scalerel*[\widthof{\ensuremath{#1}}]{\kern-.6pt\bigwedge\kern-.6pt}%
    {\rule[-\textheight/2]{1ex}{\textheight}}
  }{\textheight}%
}{0.5ex}}%
\stackon[1pt]{#1}{\tmpbox}%
}

\usepackage{acro}
\DeclareAcronym{qpu}{short=\textsc{QPU}, long=Quantum Processing Unit}
\DeclareAcronym{qsp}{short=\textsc{QSP}, long=Quantum Signal Processing}
\DeclareAcronym{qaoa}{short=\textsc{QAOA}, long=Quantum Approximate Optimization Algorithm}
\DeclareAcronym{nisq}{short=\textsc{NISQ}, long=Noisy Intermediate Scale Quantum computer}
\DeclareAcronym{ftqc}{short=\textsc{FTQC}, long=Fault-Tolerant Quantum Computer}
\DeclareAcronym{qubo}{short=\textsc{QUBO}, long=Quadratic Unconstrained Binary Optimization}
\DeclareAcronym{hubo}{short=\textsc{HUBO}, long=High-order Unconstrained Binary Optimization}
\DeclareAcronym{lcu}{short=\textsc{LCU}, long=Linear Combination of Unitaries}
\DeclareAcronym{qpe}{short=\textsc{QPE}, long=Quantum Phase Estimation}
\DeclareAcronym{vqe}{short=\textsc{VQE}, long=Variational Quantum Eigen-solver}
\DeclareAcronym{uccsd}{short=\textsc{UCCSD}, long=Unitary Coupled Cluster Simple Double}
\DeclareAcronym{mpf}{short=\textsc{MPF}, long=Multi-Product Formula}
\DeclareAcronym{qlsp}{short=\textsc{QLSP}, long=Quantum Linear System Problem}
\DeclareAcronym{scb}{short=\textsc{SCB}, long=Single Component Basis}
\DeclareAcronym{be}{short=\textsc{BE}, long=Block-encoding}
\DeclareAcronym{pde}{short=\textsc{PDE}, long=Partial Differential Equation}
\DeclareAcronym{hhl}{short=\textsc{HHL}, long=Harrow–Hassidim–Lloyd algorithm}
\DeclareAcronym{vqsl}{short=\textsc{VQSL}, long=Variational Quantum Linear Solver}

\usepackage{tikz}
\usetikzlibrary{positioning,arrows,calc,math,angles,quotes}
\usepackage{multirow}

\def\BibTeX{{\rm B\kern-.05em{\sc i\kern-.025em b}\kern-.08em
    T\kern-.1667em\lower.7ex\hbox{E}\kern-.125emX}}
\begin{document}


\title{Gate Efficient Composition of Hamiltonian Simulation and Block-Encoding \\
with its Application on \acs{hubo}, Chemistry and Finite Difference Method
}

\def\orgadep{CEA, List, F-91120}
\def\orga{CEA}
\def\loc{Palaiseau, France} 
\def\univ{Université Paris-Saclay}

\author{\IEEEauthorblockN{Robin OLLIVE}
\IEEEauthorblockA{\textit{\univ} \\
\textit{\orgadep}\\
\loc \\
0009-0006-7539-363X} 
\and
\IEEEauthorblockN{Stephane LOUISE}
\IEEEauthorblockA{\textit{\univ} \\
\textit{\orgadep}\\
\loc \\
0000-0003-4604-6453}
}

\maketitle

\begin{abstract}
    This article proposes a formalism which unifies Hamiltonian simulation techniques from different fields.
    This formalism leads to a competitive method to construct the Hamiltonian simulation with a comprehensible, simple-to-implement circuit generation technique.
    It leads to a gate decomposition and a scaling different from the usual strategy based on a \ac{lcu} reformulation of the problem.
    It can significantly reduce the quantum circuit number of rotational gates, multi-qubit gates, and the circuit depth.
    This method leads to one exact Hamiltonian simulation for each summed term and Trotter step.
    Each of these Hamiltonian simulation unitary matrices also allows the construction of the non-exponential terms with a maximum of six unitary matrices to be \acf{be}.

    The formalism is easy to apply to the widely studied \ac{hubo}, fermionic transition Hamiltonian, and basic finite difference method instances.
    For the \ac{hubo}, our implementation exponentially reduces the number of gates for high-order cost functions with respect to the \ac{hubo} order.
    The individual electronic transitions are implemented without error for the second-quantization Fermionic Hamiltonian.
    Finite difference proposed matrix decompositions are straightforward, very versatile, and scale as the state-of-the-art proposals.
\end{abstract}


\section{Introduction}
One of the major bottlenecks in quantum computing is the challenge of translating classical problems into a form that quantum computers can process.
Techniques such as Hamiltonian simulation and block encoding enable the mapping of these problems into queries suitable for quantum computation, facilitating their processing by quantum systems.
The complex exponentiation or Hamiltonian simulation can be constructed for a Hermitian matrix.
Generally, a Hermitian matrix that retains the properties can be constructed from any matrix of interest \cite{harrow_quantum_2009}.
The Hamiltonian simulation of this matrix, which is an unitary matrix (since the matrix is Hermitian) that keeps the same eigenvalues and eigenvectors, is a convenient way to process such a matrix with a \ac{qpu}.

Hamiltonian simulation called as a query is needed in many quantum computing routines.
It encompasses both \ac{nisq} and \ac{ftqc} routines such as \ac{qsp} \cite{dong_ground_2022}, \ac{qpe} \cite[p.221]{nielsen_quantum_2016}, \ac{hhl} and \ac{qaoa} \cite{farhi_quantum_2014} or \ac{vqe} \cite{peruzzo_variational_2014}.

Two main families of methods allow us to construct the Hamiltonian simulation.
The first family includes quantum-walk, \ac{lcu} or \ac{qsp} and uses as initial query the exact block encoding of the matrix of interest
\cite{martyn_grand_2021}.
These algorithms give a result with a certified error and are optimal.
However, they are resources-intensive and need many ancilla qubits and amplitude amplification.
The second family of algorithms are based on variants of the Trotter formula, also known as product formula \cite{suzuki_generalized_1976}.
These algorithms are resource-efficient but come with a Trotter error.

This paper proposes three contributions.
A simple formalism to work with this second family of algorithms is proposed.
This formalism is more adapted to treat the natural formulation of many problems.
It leads to a decomposition with often fewer terms than the state-of-the-art.
The second contribution shows how to translate a Trotter-step into a \ac{lcu}, which constructs the non-exponential terms with a maximum of six times the number of unitary matrices of a Trotter-step.
The last contribution gives examples of well-known problems.

\section{Constructing Hermitian Matrix Hamiltonian Simulation with Product Formula}
\subsection{Usual Strategy: With a Unitary Mapping}
In order to implement the Hamiltonian simulation of a matrix, the usual strategy consists of the following:
\begin{itemize}
    \item Express the problem as a Hermitian matrix:
    \begin{equation}
        \widehat{H} = \underset{i}{\sum} \alpha_{i} \widehat{H}_{i}
    \end{equation}
    \begin{itemize}
    \item The \Cref{section_qlsp} explains how to deal with non-Hermitian matrices.
    \end{itemize}
    \item Express the matrix as a \ac{lcu} \cite{koska_tree-approach_2024}: 
    \begin{equation}
    \widehat{H} = \underset{i}{\sum} \beta_{i} \widehat{PS_{i}}
    \end{equation}
    with $ \widehat{PS} = {\otimes}_{j = 0}^{N} \widehat{P}_{j} $ and $ \beta_{i} = Tr[\widehat{PS_{i}} \widehat{H}] $.
    \begin{itemize}
    \item One of the sets of unitary matrix that is the most used is, for instance, the Pauli gates plus identity $ \widehat{P} \in \{ \widehat{I}, \widehat{X}, \widehat{Y}, \widehat{Z} \} $.
    This decomposition basis is very convenient because the Hamiltonian simulation of Pauli matrices tensor product, the Pauli-strings, can be done using a linear number of 2-qubit gates \cite[p.210]{nielsen_quantum_2016}.
    In the worst case, the matrix LCU generates $4^{N}$ Pauli-strings terms (also called fragments), $N$ is the number of qubits.
    \item For some problems, among which are \ac{hubo} and chemical transitions, the technique consists of mapping directly into \ac{lcu} the terms summed to express the Hermitian matrix.
    \Cref{section_exemple} gives examples of these problems.
    \end{itemize}
    \item The last step consists of using Trotter-Suzuki's or  high-order product formula \cite[p.207]{nielsen_quantum_2016}, \cite{suzuki_generalized_1976}, cut at an order $p$ high enough to have a negligible error:
    \begin{equation}
        e^{i t \widehat{H}} = \lim\limits_{p \to \infty}\{ \prod_{i} e^{i \frac{t}{p} \beta_{i} \widehat{PS}_{i}} \}^{p}
    \end{equation}
\end{itemize}

\subsection{Direct Strategy: Without Unitary Mapping}
Instead of mapping the matrix as a \ac{lcu} and taking its Hamiltonian simulation thanks to the Trotter formula, skipping the \ac{lcu} step is possible.
It means directly expressing the complex exponential of each non-commuting term that composes the matrix of interest.
Indeed, matrix problems are often natively expressed in the \acl{scb}:
\begin{equation}
    \widehat{H} = \underset{i}{\sum} \gamma_{i} \widehat{A_{i}}
\end{equation}
with $ \widehat{A} = {\otimes}_{j = 0}^{N} \widehat{C}_{j} $ and $ \widehat{C} \in \{ \widehat{n}, \widehat{m}, \widehat{\sigma}, \widehat{\sigma}^{\dag} \} $.
The operators in this basis with the usually used mapping are expressed in \Cref{table_operateurs}.

\begin{table}[tb]
	\caption{\acl{scb} operators and their mapping in Pauli operators}
    \begin{center}
	\begin{tabular}{|ccc|}
		\hline
        \textbf{Operator} & \textbf{Matrix} & \textbf{Mapping} \\
        \hline
        $\widehat{\sigma}$ & $\begin{bmatrix}
            0 & 0 \\
            1 & 0
        \end{bmatrix} $ & $ \frac{\widehat{X} + i \widehat{Y}}{2} $ \\
        \hline
        $\widehat{\sigma}^{\dag}$ & $\begin{bmatrix}
            0 & 1 \\
            0 & 0
        \end{bmatrix} $ & $ \frac{\widehat{X} - i \widehat{Y}}{2} $ \\
        \hline
        $\widehat{n}$ & $\begin{bmatrix}
            0 & 0 \\
            0 & 1
        \end{bmatrix} $ & $ \frac{\widehat{I} - \widehat{Z}}{2} $ \\
        \hline
        $\widehat{m}$ & $\begin{bmatrix}
            1 & 0 \\
            0 & 0
        \end{bmatrix} $ & $ \frac{\widehat{I} + \widehat{Z}}{2} $ \\
        \hline
	\end{tabular}
	\end{center}
    \label{table_operateurs}
\end{table}

Instead of mapping each of these terms to a sum Pauli-strings, the terms are gathered with their Hermitian conjugate ($\mathit{h.c.}$) and then exponentiated:
\begin{equation}
    \widehat{H} = \underset{i}{\sum} (\gamma_{i} \widehat{A_{i}} + \mathit{h.c.} )
    \label{formulation_clean}
\end{equation}
\Cref{table_schema_path} illustrates the path followed by these two strategies.
The exponential of each term of \Cref{formulation_clean} can be constructed without error following the strategy described in \Cref{practice}.

\subsubsection{The Downside of Mapping}
The usual strategy mapping step leads to a number of Pauli-strings growing exponentially with the number of terms in the decomposed tensorial product (called here the term order by analogy with \ac{hubo} problems).
Because of this exponential number of terms, keeping the initial mapping the same is better, especially with high-order terms.
When applying a mapping, this problem of an exponential number of terms also appears when Pauli-strings are mapped to \acl{scb}.
In order to avoid it, the two strategies can be mixed.
The method to construct a mixed quantum circuit is described in \Cref{practice}.

\begin{figure}[tb]
    \begin{center}
	\begin{tabular}{ccc}
		$ \widehat{H} = \underset{j}{\sum} \gamma_{j} \widehat{A_{j}} $ & $\longrightarrow$ & $ \widehat{H} = \underset{i}{\sum} \beta_{i} \widehat{PS_{i}} $ \\
        $\downarrow$ & & $\downarrow$ \\
        $ e^{i t \widehat{H}} \simeq \prod_{j} e^{i t \gamma_{j} \widehat{A}_{j}} $ & & $ e^{i t \widehat{H}} \simeq \prod_{i} e^{i t \beta_{i} \widehat{PS}_{i}} $ \\
        $ \simeq \prod_{j} \widehat{V}_{j} $ & $\neq$ & $ \simeq \prod_{i} \widehat{U}_{i} $ \\
        & & \\
        \textbf{Direct} & & \textbf{Usual}
	\end{tabular}
	\caption{Schematic of the two Hamiltonian simulation strategies path (at order $p = 1$)}
    \label{table_schema_path}
	\end{center}
\end{figure}

\section{Direct Hamiltonian Simulation in Practice}
\label{practice}
\begin{figure*}[tb]
\begin{center}
\resizebox{\linewidth}{!}{\includegraphics{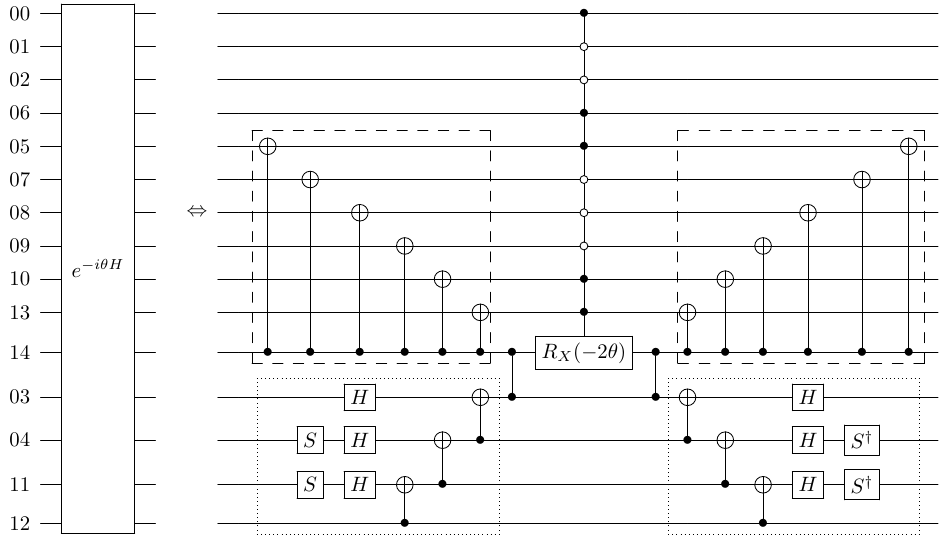}}
\end{center}
\caption{$\exp(- i \theta \widehat{H})$ with $ \widehat{H} = \widehat{n}_{0} \widehat{m}_{1} \widehat{m}_{2} \widehat{X}_{3} \widehat{Y}_{4} \widehat{\sigma}_{5}^{\dag} \widehat{n}_{6} \widehat{\sigma}_{7} \widehat{\sigma}_{8} \widehat{\sigma}_{9} \widehat{\sigma}_{10}^{\dag} \widehat{Y}_{11} \widehat{Z}_{12} \widehat{\sigma}_{13}^{\dag} \widehat{\sigma}_{14} + \mathit{h.c.} $ gate decomposition, the number on the left of the qubit contains the qubit index.}
\label{full_circuit_texte}
\end{figure*}

This section describes how to exponentiate an arbitrary term composed as a tensorial product of $\{ \widehat{I}, \widehat{X}, \widehat{Y}, \widehat{Z}, \widehat{n}, \widehat{m}, \widehat{\sigma}, \widehat{\sigma}^{\dag} \}$. To achieve that, the Hamiltonian simulation of $$ \widehat{H} = \widehat{n} \widehat{m} \widehat{m} \widehat{X} \widehat{Y} \widehat{\sigma}^{\dag} \widehat{n} \widehat{\sigma} \widehat{\sigma} \widehat{\sigma} \widehat{\sigma}^{\dag} \widehat{Y} \widehat{Z} \widehat{\sigma}^{\dag} \widehat{\sigma} + \mathit{h.c.} $$ is constructed as an example in \Cref{full_circuit_texte}.
The usual mapping expresses this unitary operator sum of $ 2^{11} = 2048 $ Pauli-strings.

\begin{itemize}
\item First, gather (virtually) the operator sub-matrices in four families which are treated differently:
\begin{itemize}
\item The identity matrix $\widehat{I}$
\item The Pauli $ \{ \widehat{X}, \widehat{Y}, \widehat{Z} \} $
\item The number of exitations or control operators $ \{ \widehat{n}, \widehat{m} \} $
\item The excitation or transition operators $ \{ \widehat{\sigma}, \widehat{\sigma}^{\dag} \} $
\end{itemize}
\item The first family contains only the identity operator for which no action is needed on the qubit associated with the corresponding indexes.
\item The second family is the transition operators:
\begin{itemize}
\item A basis change toward the basis in which the two states of the transition are diagonal.
In this basis, the two $n + 1$ qubit states can be identified by the same $n$-controlled-gate.
Since going from one state to the other is done only by switching the zero and one of the state indexes in binary format, equivalently, the states are each other ones' complement; it is only needed to check that the bit contains the same value on the $\sigma$ index and the other value on the $\sigma^{\dag}$ index without knowing this value:
\begin{equation}
    \begin{aligned}
        \widehat{H}_{\sigma} & = \prod \widehat{\sigma} \prod \widehat{\sigma}^{\dag} + \mathit{h.c.} \\
        & = \prod \ket{0} \bra{1} \prod\ket{1} \bra{0} + \mathit{h.c.} \\
        & = \ket{\mathrm{bin}[q]} \bra{\mathrm{bin}[-q]} + \mathit{h.c.}
    \end{aligned}
\end{equation}
This state corresponds to a generalized Bell state, and going to the bell state basis can easily be done thanks to parity gates and a Hadamard gate hidden in the rotational gate.
It is illustrated by the dashed box of the \Cref{full_circuit_texte}.
\item A rotation can then be done between the two states before uncomputing the circuit.
\end{itemize}
\item For the Pauli-string, the whole string possesses the two $2^{n-1}$ ($n$ the number of qubits of the Pauli-string) degenerate eigenvalues which are one and minus-one:
\begin{equation}
    \widehat{H} = \widehat{PS} \widehat{H}_{n \sigma} = \widehat{W}
    \begin{bmatrix}
        \widehat{H}_{n \sigma} & 0 \\
        0 & - \widehat{H}_{n \sigma}
    \end{bmatrix} \widehat{W}^{\dag}
\end{equation}
with $\widehat{W}^{\dag}$ the basis change between the computationnal and $\widehat{PS}$ eigenbasis, since $\widehat{H}_{n \sigma}$ and $\widehat{W}$ apply to different qubit groups, they commute.
When the operator is exponentiated, the plus or minus-one controls the sign of the rotation.
The usual technique to check their eigenbasis is:
\begin{itemize}
\item Go in the individual operator basis thanks to $ \{ \widehat{H}, \widehat{S}, \widehat{S}^{\dag} \} $ gates.
\item Report the parity of all the Pauli on the same qubit, illustrated by the gates bordered by dots.
\item Use this qubit to control the sign of the phases.
\item The sign is controlled by taking advantage of: $ \widehat{R_{X/Y}}(\theta) = \widehat{Z} \widehat{R_{X/Y}}(- \theta) \widehat{Z} $.
\item Uncompute the basis change.
\end{itemize}
\item For the number operator family, it is easy to understand that they become control of the key represented by the associated states of the rest of the gate when exponentiated: $ \widehat{H} = e^{i \widehat{n} \widehat{m} \widehat{m} \widehat{n} \widehat{H}_{PS \sigma}} = \reallywidehat{C^{4}\{\ket{1001}\} e^{i H_{PS \sigma}}} $.
\end{itemize}

\subsection{Complex Component}
Hermitian matrix can possess out-of-diagonal complex components.
In order to deal with such a component, the previous process can be used to go in the component basis, but in order to implement both the real and imaginary part, the operator must be split:
\begin{equation}
\begin{aligned}
    \widehat{H}_{comp} & = z \left( \prod \widehat{\sigma} \prod \widehat{\sigma}^{\dag} \right) \widehat{H}_{n PS} + \mathit{h.c.}\\
    &  = \left( z \prod \widehat{\sigma} \prod \widehat{\sigma}^{\dag} + \mathit{h.c.} \right) \widehat{H}_{n PS} \\
    & = \mathrm{Re}[z] \widehat{H} + i \mathrm{Im}[z] \left( \prod \widehat{\sigma} \prod \widehat{\sigma}^{\dag} - \mathit{h.c.} \right) \widehat{H}_{n PS}
\end{aligned}
\end{equation}
The \Cref{full_circuit_texte} need to be modified by changing:
\begin{equation*}
    \widehat{R_{X}}(-2 \theta) \leftarrow \widehat{R_{X}}(-2 \mathrm{Re}[z] \theta) \widehat{R_{Y}}(-2 \mathrm{Im}[z] \theta)
\end{equation*}
Since these two operators do not commute, a Trotter error appears.

If the complex coefficient is on the diagonal, the matrix is no longer Hermitian, so the technique proposed in the \Cref{section_qlsp} must be used.

\subsection{Parity Gate Depth Optimization}
The naive basis change proposed at the beginning of the section to go from the computational basis to the transition operator eigenbasis is very interesting:
This basis change gives access to the anti-diagonal of the sub-matrix composed by the transition operators:
\begin{equation}
    \ket{1} \bra{0} \bigotimes ( \ket{\mathrm{bin}[q]} \bra{\mathrm{bin}[-q]} + \mathit{h.c.} )
\end{equation}
In the case that the Hamiltonian contains another term that can be written as $q \in \mathbb{N}$, it is only required to apply the corresponding controlled-rotational gate instead of the whole basis change.

\begin{figure}[tb]
\begin{center}
\resizebox{\linewidth}{!}{
\includegraphics{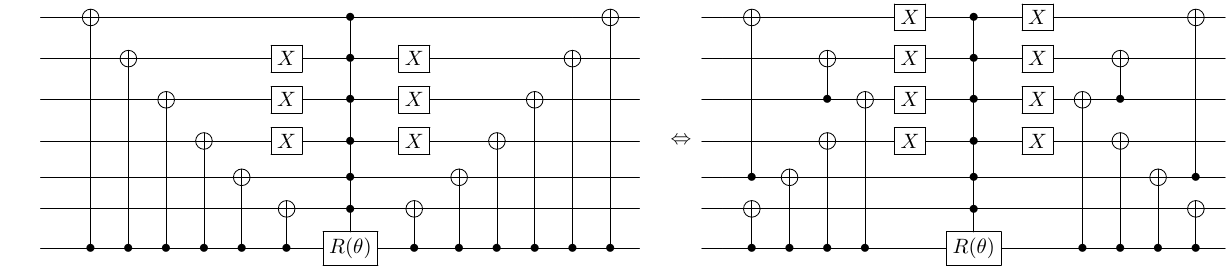}}
\end{center}
\caption{These two basis changes allow to turn around $ \ket{1000110} \bra{0111001} + \mathit{h.c.} $ in the qubitized Bloch sphere.}
\label{short_basis_change}
\end{figure}

If this property cannot compact different gates, another basis change leading to a shallower circuit can be used, as illustrated by \Cref{short_basis_change}.
Instead of being linear in the number of two-qubit gates and linear in-depth, this other basis change is sub-linear in deepth while preserving the same number of two-qubit gates.
This basis change checks the qubits that must have the same value thanks to a pyramidal structure two-by-two, so they do not always use the same qubit.

\section{Constructing the \acl{lcu} in Practice}
This section describes how to construct efficiently the \ac{lcu} corresponding to one non-exponential term of the Hamiltonian from its associated Hamiltonian simulation unitary matrix.

Using the same four families as in the previous \Cref{practice}, the Pauli and identity matrices are already unitaries, so only the remaining families must be mapped as unitary matrices.

\begin{itemize}
\item The whole control family can be mapped as a sum of two unitary matrices.
To achieve that, it needs:
\begin{itemize}
\item an identity matrix
\item an $n$-controlled -Z-gate acting on the state controlled by the key of the number operators: $$ \widehat{C^{n}Z}\{ \ket{a} \} = \widehat{I} - 2 \ket{a} \bra{a} $$
\item The unitary mapping is: 
\begin{equation} 
    \widehat{H}_{n} = \ket{a} \bra{a} = \frac{\widehat{I} - \widehat{C^{n}Z}\{ \ket{a} \}}{2}
\end{equation}
\end{itemize}
\item The whole transition family can be mapped as a sum of three unitary matrices.
\begin{itemize}
    \item an identity matrix
    \item two consecutive $n$-controlled-Z-gate acting on the state controlled by the key of the number operators:
    \begin{equation*}
    \begin{aligned}
        & \widehat{C^{n}Z}\widehat{C^{n}Z}\{ \ket{a}; \ket{b} \} = \widehat{I} - 2 (\ket{a} \bra{a} + \ket{b} \bra{b}) \\
        \Leftrightarrow & \frac{\widehat{I} + \widehat{C^{n}Z}\widehat{C^{n}Z}\{ \ket{a}; \ket{b} \}}{2} = \widehat{I} - (\ket{a} \bra{a} + \ket{b} \bra{b})
    \end{aligned}
    \end{equation*}
    \item The modified \Cref{full_circuit_texte} gate:
    \begin{equation*}
        \widehat{R_{X}}(-2 \theta) \leftarrow \widehat{X}
    \end{equation*}
    This gate corresponds to the operator $\widehat{C^{n}X}\{ \ket{a}; \ket{b} \}$.
    \item The mapping is:
    \begin{equation}
    \begin{aligned}
        & \widehat{C^{n}X}\{ \ket{a}; \ket{b} \} = \widehat{H}_{\sigma} + \widehat{I} - (\ket{a} \bra{a} + \ket{b} \bra{b}) \\
        \Leftrightarrow & \widehat{H}_{\sigma} = \widehat{C^{n}X}\{ \ket{a}; \ket{b} \} + \frac{\widehat{I} + \widehat{C^{n}Z}\widehat{C^{n}Z}\{ \ket{a}; \ket{b} \}}{2}
    \end{aligned}
    \end{equation}
    \end{itemize}
\end{itemize}

If both mappings are needed, the \ac{be} implementation is done with at most six unitary matrices for each Hamiltonian simulation term:

{\small
\begin{equation}
\begin{aligned}
    \widehat{H} & = \widehat{n}_{0} \widehat{m}_{1} \widehat{m}_{2} \widehat{X}_{3} \widehat{Y}_{4} \widehat{\sigma}_{5}^{\dag} \widehat{n}_{6} \widehat{\sigma}_{7} \widehat{\sigma}_{8} \widehat{\sigma}_{9} \widehat{\sigma}_{10}^{\dag} \widehat{Y}_{11} \widehat{Z}_{12} \widehat{\sigma}_{13}^{\dag} \widehat{\sigma}_{14} + \mathit{h.c.} \\
    & = (\widehat{\sigma}_{5}^{\dag} \widehat{\sigma}_{7} \widehat{\sigma}_{8} \widehat{\sigma}_{9} \widehat{\sigma}_{10}^{\dag} \widehat{\sigma}_{13}^{\dag} \widehat{\sigma}_{14} + \mathit{h.c.}) \bigotimes  \widehat{n}_{0} \widehat{m}_{1} \widehat{m}_{2} \widehat{n}_{6} \bigotimes \widehat{X}_{3} \widehat{Y}_{4} \widehat{Y}_{11} \widehat{Z}_{12} \\
    & = \widehat{H}_{\sigma} \bigotimes \widehat{H}_{n} \bigotimes \widehat{H}_{PS} \\
    & = \left( \widehat{C^{6}X}\{ \ket{a}; \ket{b} \} + \frac{\widehat{I} + \widehat{C^{6}Z}\widehat{C^{6}Z}\{ \ket{a}; \ket{b} \}}{2} \right) \\
    & \qquad \qquad \qquad \qquad \qquad \bigotimes \frac{\widehat{I} - \widehat{C^{3}Z}\{ \ket{c} \}}{2} \bigotimes \widehat{X}_{3} \widehat{Y}_{4} \widehat{Y}_{11} \widehat{Z}_{12}
\end{aligned}
\end{equation}
}
with
\begin{itemize}
\item $ \ket{a} = \ket{0 1 1 1 0 0 1}_{5, 7, 8, 9, 10, 13, 14} $
\item $ \ket{b} = \ket{1 0 0 0 1 1 0}_{5, 7, 8, 9, 10, 13, 14} $
\item $ \ket{c} = \ket{1001}_{0, 1, 2, 6} $
\end{itemize}

\begin{figure}[tb]
\begin{center}
\resizebox{\linewidth}{!}{
\includegraphics{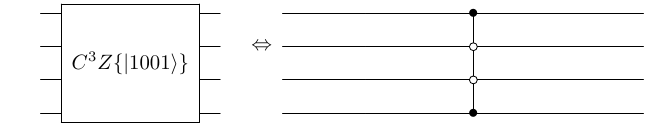}}
\end{center}
\caption{$ \widehat{C^{3}Z}\{ \ket{1001} \} $ gate construction.}
\label{gate_lcu1}
\end{figure}

\begin{figure}[tb]
\begin{center}
\resizebox{\linewidth}{!}{
\includegraphics{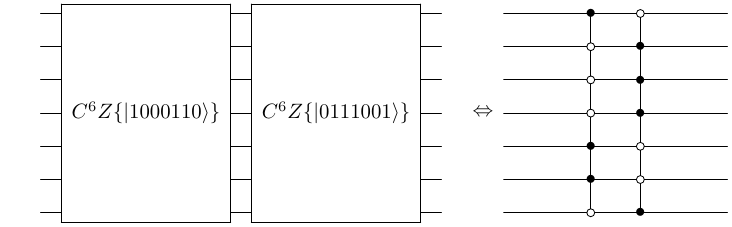}}
\end{center}
\caption{$ \widehat{C^{6}Z}\widehat{C^{6}Z}\{ \ket{1000110}; \ket{0111001} \} $ gate construction.}
\label{gate_lcu2}
\end{figure}

\begin{figure}[tb]
\begin{center}
\resizebox{\linewidth}{!}{
\includegraphics{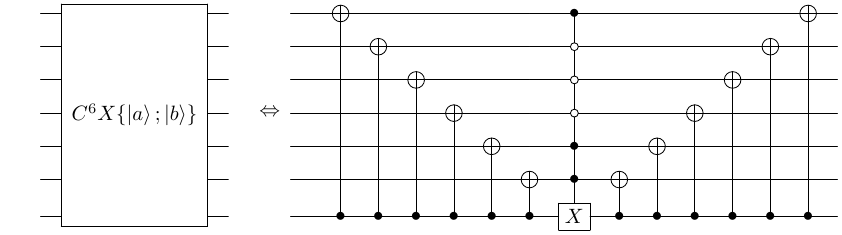}}
\end{center}
\caption{$ \widehat{C^{6}X}\{ \ket{1000110}; \ket{0111001} \} $ gate construction.}
\label{gate_lcu3}
\end{figure}

\section{Examples}
\label{section_exemple}
\subsection{\acl{hubo}}
Some problems, such as hypergraph max-cut, are expressed with a unitary formalism:
\begin{equation}
    \widehat{H_{P}} = \sum_{\mathcal{I} \in \mathbb{P}(\{ 1, ..., n \})} \biggl[ q_{\mathcal{I}} \prod_{i \in \mathcal{I}} \widehat{Z_{i}} \biggr]
    \label{hubo_eq}
\end{equation}
with the problem weights, $ q_{\mathcal{I}} \in \mathbb{R} $, $ \mathcal{I} \in \mathbb{P}( \{ 1, ..., n \} ) $, $\mathbb{P}(\mathbb{E})$ is a powerset of $\mathbb{E}$.

The Ising or weighted max-cut instances are a specific case of order two, equivalent to the \ac{qubo} problem \cite{lucas_ising_2014}.
Many other \ac{hubo} problems, such as the knapsack problem, are often expressed or reduced to the following boolean-formalism using the number of excitation operators:
\begin{equation}
    \widehat{H_{P}} = \sum_{\mathcal{I} \in \mathbb{P}(\{ 1, ..., n \})} \biggl[ b_{\mathcal{I}} \prod_{i \in \mathcal{I}} \widehat{n}_{i} \biggr]
    \label{hubo_nonunitary_eq}
\end{equation}

In both strategies, there is no trotter error since the gates commute.
The Pauli-strings and the single-component techniques can thus be easily compared for optimization problems.
It is worth mentioning that:
\begin{itemize}
    \item both the $\widehat{R_{Z^{n}}}$ and the $\widehat{C^{n}P}$ gates can be decomposed in two-qubit gates with a linear two-qubits gate number $m$ \cite[Sec.7: A, B and E]{barenco_elementary_1995} \cite[p.182]{nielsen_quantum_2016}.
    \item The two-qubit gates number is $ m = 2 (n - 1) $ for $\widehat{R_{Z^{n}}}$ and $ m = 2 \times ( 6 \times 8 (n - 5) + 48 n - 212 ) \propto 192 n $ plus an ancilla qubit if $n > 5$ for $\widehat{C^{n}P}$.
    \item $\widehat{C^{n}P}$ can also be decomposed without ancilla qubits in two-qubit gates with a quadratic cost with respect to the number of controls.
    \item For a dense problem, the number of terms will stay the same for the two formalisms.
    \item For sparse matrices, switching of formalism transforms each term in a sum of $ 2^{n} - 1 $ terms with $n$ the order of the term.
\end{itemize}

The last point suggests that for sparse problems of high order, staying in the initial formalism leads directly to a circuit exponentially shallower (with respect to the higher order of the problem variables) while not adding an extra qubit to encode the problem\footnote{
Ising quadratization can reduce this advantage at the cost of higher problem size and extra classical computations.}.
For a \ac{qpu} that can execute $ \{ \widehat{R_{Z}}, \widehat{CX}, \widehat{P}, \widehat{CP} \} $, it is possible to show\footnote{
    The number of states of each order $h$ that compose a gate of order $n$ when the formulation is switched is the combinatorics of $h$ in $n$:

$$ 2^{n} - 1 = \sum_{h = 1}^{n} C_{n}^{h} $$

The following equation is equivalent to seeking the 2-qubit-gate number by the direct method inferior to the number of 2-qubit gates by the usual method for a dense problem with the maximum order $n > 5$:
\begin{equation*}
\begin{aligned}
    2 (6 \times 8 (n - 5) + 48 n - 212) & < \sum_{h = 1}^{n} 2 (h - 1) C_{n}^{h} \\
    n > 7
\end{aligned}
\end{equation*}
} that the $\widehat{C^{n}P}$ decomposition in two-qubit gates leads to shallower circuits with the direct method when $n > 7$.
If the $\widehat{CP}$ gate is not part of the gate set, the decomposition is also advantageous but with a higher threshold and requires a small overhead in the number of gates.
The usual method is the most efficient if the problem has a similar density in the two formalisms.

Theoretically, a 'perfect' compiler would go to the optimal result, whatever the formalism used, but unfortunately, there is no recipe for a 'perfect' compiler without an exponential cost yet.

\subsubsection{Origin of the Direct Strategy Idea}
The main idea of the direct strategy came from an article that constructs Hamiltonian simulation with the gates of the direct strategy for a \ac{qubo} problem.
They construct this Hamiltonian simulation indirectly without linking their strategy to a \ac{qpe}.
It proposes constructing a quantum circuit that allows reading a superposition of the eigenstate of a \ac{hubo} problem.
While this method can be identified as a \ac{qpe}, the problem Hamiltonian was not decomposed with the usual Pauli strings \cite[Fig. 4, 5 \& 6]{gilliam_grover_2021}.
The direct Hamiltonian simulation strategy explains the difference between the expected and identified Hamiltonian.

\subsection{Fermionic Hamiltonian}
Electronic (fermionic) Hamiltonians can be expressed as:
\begin{equation}
    \widehat{H}_{ferm} = \sum_{i j} h_{i j} \widehat{a}_{i}^{\dag} \widehat{a}_{j} + \sum_{i j k l} h_{i j k l} \widehat{a}_{i}^{\dag} \widehat{a}_{j}^{\dag} \widehat{a}_{k} \widehat{a}_{l}
\end{equation}
With the fermionic ladders operator that keeps the fermionic anti-symmetries.
These ladders operators can be expressed with the Jordan-Winger mappings\footnote{Other mappings, such as the Bravyi-Kitaev or parity mapping, exist to switch from fermionic to qubit ladders operators \cite{seeley_bravyi-kitaev_2012}. While more efficient than the other mappings, the Bravyi-Kitaev is still based on \ac{lcu} mapping before the Trotterisation.}\cite{jordan_uber_1928}: $$ \widehat{a}_{i} = \widehat{\sigma}_{i} \prod_{j = 0}^{i - 1} \widehat{Z}_{j} $$

Applying the direct strategy at this stage is impossible since $\widehat{\sigma}$ is not Hermitian.
The ladder operators can be gathered into Hermitian matrices:
\begin{equation}
\begin{aligned}
    \widehat{H}_{ferm} & = \sum_{i j} \frac{h_{i j}}{2} ( \widehat{a}_{i}^{\dag} \widehat{a}_{j} + \mathit{h.c.} ) + \sum_{i j k l} \frac{h_{i j k l}}{2} ( \widehat{a}_{i}^{\dag} \widehat{a}_{j}^{\dag} \widehat{a}_{k} \widehat{a}_{l} + \mathit{h.c.} ) \\
    & = \sum_{i j} \frac{h_{i j}}{2} \widehat{H}_{1 i j} + \sum_{i j k l} \frac{h_{i j k l}}{2} \widehat{H}_{2 ijkl}
\end{aligned}
\end{equation}

The following discussion concerns one-body terms but can be extended to two-body terms.
When applying the fermionic anti-symmetry, it leads to:
\begin{equation}
\begin{aligned}
    \widehat{H}_{1 i j} & = ( \widehat{\sigma}_{i}^{\dag} \widehat{\sigma}_{j} + \mathit{h.c.}) \prod_{k = i + 1}^{j - 1} \widehat{Z}_{k} \\
    & = \widehat{A}_{1 i j} \prod_{k = i + 1}^{j - 1} \widehat{Z}_{k}
\end{aligned}
\end{equation}

\subsubsection{Implementation, Individual Electronic Transition}
As presented in the \Cref{practice}, a mix of Pauli-strings and \acl{scb} can also be exponentiated; it leads to a structure similar to the tables of \cite[Table A1]{whitfield_simulation_2011}.

The one-body term gate is discussed in works about Fermi-Hubbard Hamiltonian \cite[Appendix J]{babbush_low-depth_2018} and to change of fermionic basis \cite{martinez-martinez_assessment_2023}, two-body gate are derived with another strategy by \cite{mukhopadhyay_synthesizing_2023}.

\subsubsection{Implementation, full Hamiltonian Simulation}
An extra Trotter error will appear since the different unitary matrices found with the two strategies do not commute.
The correction by circuit repetition will increase the circuit depth differently for the two strategies since the Trotter errors differ.
Different ways of gathering the term depending on their commutation relations will add an extra complexity layer to the efficient method analysis \cite{childs_theory_2021}.
\cite{martinez-martinez_assessment_2023} experimental benchmark indicates that fermionic partitioning leads to less Trotter error.

\subsubsection{Meaning of the Different Terms}
It is also interesting to note that this structure can be easily interpreted as a succession of electronic transitions while the meaning of the different unitary gates decomposition Hamiltonian simulation is somewhat lost when an \ac{lcu} decomposition is used.
The individual transitions have thus no Trotterization error, and ansatz such as \ac{uccsd} thus mimic a series of electronic transitions without error.

\subsubsection{Other Alternatives to LCU Hamiltonian Simulation}
The fermionic partitioning method\cite{martinez-martinez_assessment_2023} (also referred to as Split-operator technique \cite[Appendix C]{kivlichan_improved_2020}) also models electronic transition without relying on the unitary mapping Hamiltonian simulation.
They both classically diagonalize some gathered fragments to implement them on a diagonal basis.
In general, it leads to a complex circuit that needs a polynomial number of $e^{i \theta_{j} \widehat{A}_{1}}$ gates (with the $\theta_{j}$ parameters computed classically) to do the basis change for each term and each Trotter step \cite[Sect. 2.3 \& 3.2]{martinez-martinez_assessment_2023}.
$e^{i \theta \widehat{A}_{1}}$ is referred to as orbital rotation
transformations in these papers.

Fermionic swap strategy \cite[Appendix B]{kivlichan_improved_2020} consists in using the fermionic swap operator FSWAP to change the qubit indexes to avoid the long controlled parity gates, and so reduce $e^{i \theta_{j} \widehat{H}_{1}}$ to $e^{i \theta_{j} \widehat{A}_{1}}$.
Even if the FSWAP gates have a higher cost than the parity control gate, it allows the construction of an efficient basis change toward momentum space \cite{kivlichan_improved_2020} under some periodic conditions of the system.

\subsection{Finite Difference Methode}
Solving \ac{pde} is another significant computing problem.
General \ac{pde} are very complex to solve but can always be linearised at high-order approximation.
Many physics problems are formulated thanks to linear \ac{pde}.
Linear \ac{pde} systems are systems of equation and so can be expressed as matrix problems:
\begin{equation}
\begin{aligned}
    \beta & = \left( \alpha + \alpha_{x 1} \frac{\partial}{\partial x} + \alpha_{y 1} \frac{\partial}{\partial y} + \alpha_{x 2} \frac{\partial^{2}}{\partial x^{2}} + \alpha_{y 2} \frac{\partial^{2}}{\partial y^{2}} \right) f(x, y) \\
    & = \begin{bmatrix}
        \alpha & \alpha_{x 1} & \alpha_{y 1} & \alpha_{x 2} & \alpha_{y 2}
    \end{bmatrix} . \begin{bmatrix}
        1 \\
        \frac{\partial}{\partial x} \\
        \frac{\partial}{\partial y} \\
        \frac{\partial^{2}}{\partial x^{2}} \\
        \frac{\partial^{2}}{\partial y^{2}}
    \end{bmatrix} f(x, y) \\
    & = \underline{\alpha} . \begin{bmatrix}
        1 &
        \frac{\partial}{\partial x} &
        \frac{\partial}{\partial y} &
        \frac{\partial^{2}}{\partial x^{2}} &
        \frac{\partial^{2}}{\partial y^{2}}
    \end{bmatrix}^{T} f(x, y)
\end{aligned}
\end{equation}
for a scalar function of order two without crossed-term, if $f$ were a vectorial field, then $\underline{\alpha}(x, y)$ would become a matrix (spatial dependences) and $\beta(x, y)$ a vector containing all the second members (boundary conditions).

Unfortunately, the boundary conditions, which can be complex space functions, do not allow the direct solving of this problem in general.
Many matrix formulation methods were developed to solve linear \ac{pde}.
These methods often aim to search a problem specific solution approximation.
One of the most common strategies is to use 
a space discretization, called a mesh, to express the problem locally.
It leads to methods such as finite differences, finite elements, or finite volumes.
These methods are currently state-of-the-art in finding solutions to many problems in classical computing.
Strategy to implement them in quantum computers emerged \cite[Related Works sub-section]{ty_double-logarithmic_2024}.
Algorithm such as \ac{hhl}, \ac{qsp}, or \ac{vqsl} are used to invert the matrix.

This sub-section shows that the techniques presented in this paper lead directly to competitive matrix-problem quantum implementation.
As an example, the following simple instances are studied:
\begin{itemize}
\item the simplest space discretization method: the finite difference.
\item the Poisson equation as the solved \ac{pde}.
\item the regular cartesian grid is used as the mesh in:
\begin{itemize}
\item one dimension: equidistant nodes on a line
\item two dimensions, two node-lines: nodes at the summit of square cells
\item three dimensions, two layers of two lines: nodes at the summit of cubes
\end{itemize}
\end{itemize}

\begin{figure}[tb]
\begin{center}
\resizebox{\linewidth}{!}{\includegraphics{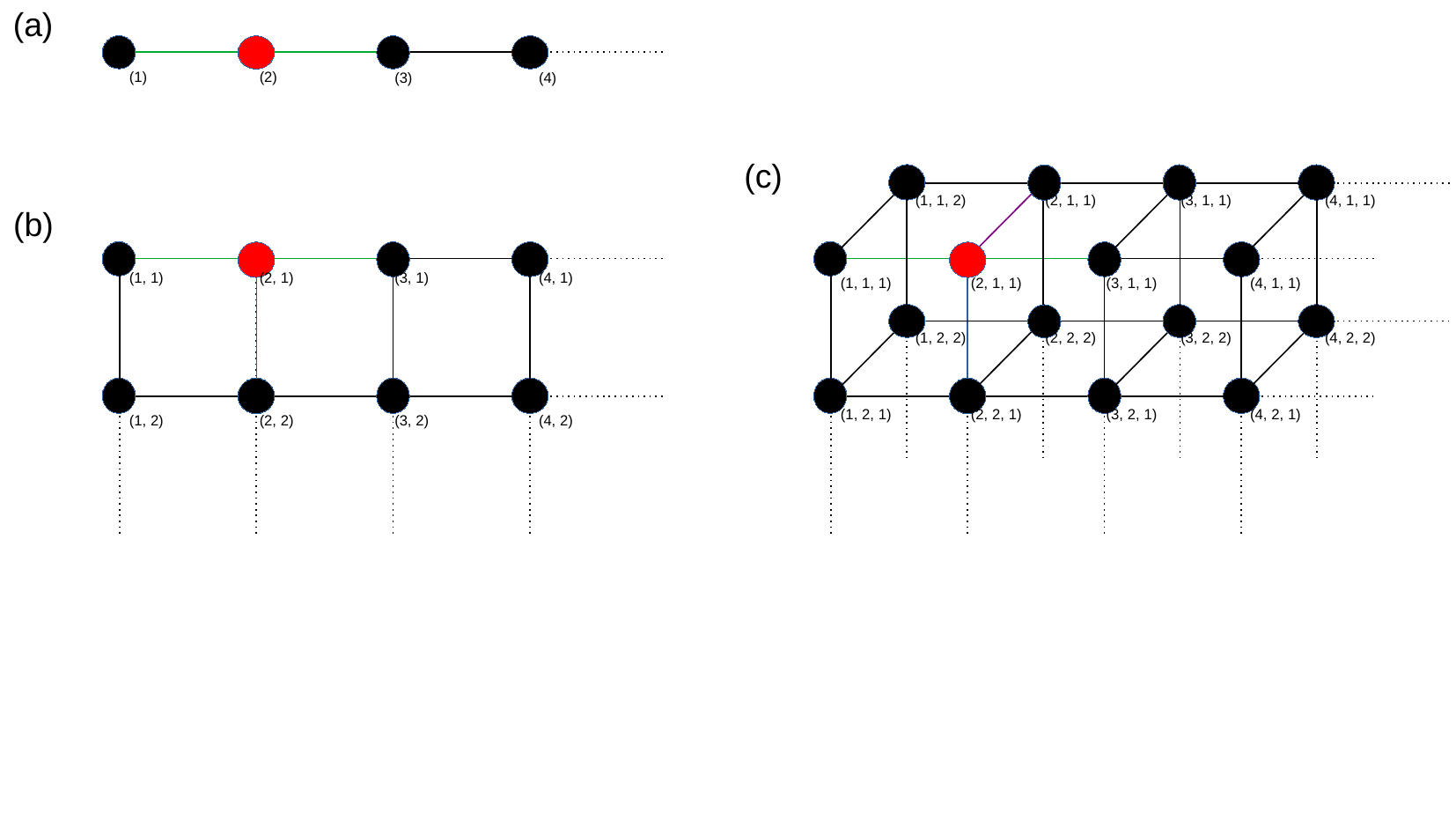}}
\end{center}
\vspace{-1.5 cm}
\caption{Schematics of the different space discretizations studied. (a) in one dimension line, (b) a two dimensions layer, and (c) three dimensions double layers.}
\label{finite_grid}
\end{figure}

\subsubsection{Reminder: Formulation of the Vectorial Analysis Operator on a Cartesian Grid}
Using the Taylor formula, the function can be derived on a point plus an infinitesimal distance:
\begin{equation}
    \begin{aligned}
\frac{\partial}{\partial x} f & \simeq \frac{f(x + d) - f(x - d)}{2 d} \\
\frac{\partial^{2}}{\partial x^{2}} f & \simeq \frac{f(x + d) + f(x - d) - 2 f(x)}{d^{2}} 
\end{aligned}
\end{equation}
this infinitesimal distance can then be changed by the distance to the adjacent nodes so that the derivative is expressed by the functions associated with its first neighbors:
\begin{equation}
\begin{aligned}
\frac{\partial}{\partial x} f_{i} & \simeq \frac{f_{i + 1} - f_{i - 1}}{2 d} \\
\frac{\partial^{2}}{\partial x^{2}} f_{i} & \simeq \frac{f_{i + 1} + f_{i - 1} - 2 f_{i}}{d^{2}}
\end{aligned}
\end{equation}
here for the grid proposed below.
The notation $f_{i j k}$ expresses the value of $f$ at the node $i, j, k$, where $i$ is the node index on the line, $j$ the index of the line, and $k$ the index of the layer of lines.
The derivative can then be assembled to recompose the local vectorial analysis operators:
\begin{equation}
\begin{aligned}
\Delta f_{i j} & \simeq \frac{f_{i + 1, j} + f_{i - 1, j} + f_{i, j + 1} + f_{i, j - 1} - 4 f_{i j}}{d^{2}} \\
\Delta f_{i j k} & \simeq \frac{f_{i + 1, j, k} + f_{i - 1, j, k} + f_{i, j + 1, k} + f_{i, j - 1, k}}{d^{2}} \\
& + \frac{f_{i, j, k + 1} + f_{i, j, k - 1} - 6 f_{i j k}}{d^{2}}
\end{aligned}
\end{equation}
and be reinjected into the equation of interest\footnote{While this paper focussed on static problems, the finite difference framework also allows the treatment of dynamic equations.
Time can be derived as a fourth index: $ f_{i, j, k, \delta t} $.
In simple explicit sheme, the time operator local approximation is expressed with respect to the nearest spatial neighbors and of the previous time neighbors $f_{\delta t - 1}$.
The problem can be solved time-step by time-step thanks to matrix product.

In a simple implicit sheme, the time operator local approximation is expressed with respect to the nearest spatial neighbors and the next time neighbors.
The problem can be solved thanks to matrix reversal.
If the sources are coupled to the system evolution, the time evolution must also be discretized in time in order to update them.}.

\subsubsection{Easy to Implement Quantum Operator Decomposition}
The Poisson equation's \ac{pde}:
\begin{equation}
\begin{aligned}
0 & = \alpha \Delta f(x, y, z) \\
\underline{0} & \simeq \alpha \underline{\sum^{i, j, k} \Delta f_{i j k}} \\
& \simeq \alpha \underline{\underline{A}} . \begin{bmatrix}
    f_{1 1 1} \\
    \vdots \\
    f_{N, 1, 1} \\
    f_{1, 2, 1} \\
    \vdots \\
    f_{N, 2, 1} \\
    \vdots \\
    f_{N, N, N}
\end{bmatrix}
\end{aligned}
\label{eq_pde_simple}
\end{equation}
with $N$, the number of nodes per line, is enough to show the matrix shape due to the first neighbor-only interaction.
If the \ac{pde} contains other vectorial analysis operators, the same components are used but with different values.
Matrices addressing individually each node-lines are convenient to address a broader range of other \ac{pde}:
\begin{itemize}
\item one dimension:
\begin{equation*}
    \begin{aligned}
\underline{\underline{A}} & = \begin{bmatrix}
    {\color{red} a_{1}} & {\color{green} a_{i 1}} & 0 & 0 \\
    {\color{green} a_{i 1}} & {\color{red} a_{1}} & {\color{green} a_{i 1}} & 0 \\
    0 & {\color{green} a_{i 1}} & {\color{red} a_{1}} & {\color{green} a_{i 1}} \\
    0 & 0 & {\color{green} a_{i 1}} & {\color{red} a_{1}}
\end{bmatrix} \\
& \Leftrightarrow a_{1} \widehat{I} \widehat{I} + a_{i 1} (\widehat{I} \widehat{X} + (\widehat{\sigma}^{\dag} \widehat{\sigma} + \mathit{h.c.}) )
\end{aligned}
\end{equation*}
\item two dimensions:
\begin{equation*}
    \begin{aligned}
\underline{\underline{A}} & = \begin{bmatrix}
    {\color{red} a_{1}} & {\color{green} a_{i 1}} & 0 & 0 & {\color{blue} a_{j 1 2}} & 0 & 0 & 0 \\
    {\color{green} a_{i 1}} & {\color{red} a_{1}} & {\color{green} a_{i 1}} & 0 & 0 & {\color{blue} a_{j 1 2}} & 0 & 0 \\
    0 & {\color{green} a_{i 1}} & {\color{red} a_{1}} & {\color{green} a_{i 1}} & 0 & 0 & {\color{blue} a_{j 1 2}} & 0 \\
    0 & 0 & {\color{green} a_{i 1}} & {\color{red} a_{1}} & 0 & 0 & 0 & {\color{blue} a_{j 1 2}} \\
    {\color{blue} a_{j 1 2}} & 0 & 0 & 0 & {\color{orange} a_{2}} & {\color{lime} a_{i 2}} & 0 & 0 \\
    0 & {\color{blue} a_{j 1 2}} & 0 & 0 & {\color{lime} a_{i 2}} & {\color{orange} a_{2}} & {\color{lime} a_{i 2}} & 0 \\
    0 & 0 & {\color{blue} a_{j 1 2}} & 0 & 0 & {\color{lime} a_{i 2}} & {\color{orange} a_{2}} & {\color{lime} a_{i 2}} \\
    0 & 0 & 0 & {\color{blue} a_{j 1 2}} & 0 & 0 & {\color{lime} a_{i 2}} & {\color{orange} a_{2}}
\end{bmatrix} \\
& \Leftrightarrow \widehat{m} (a_{1} \widehat{I} \widehat{I} + a_{i 1} (\widehat{I} \widehat{X} + (\widehat{\sigma}^{\dag} \widehat{\sigma} + \mathit{h.c.}) )) \\
& + \widehat{n} (a_{2} \widehat{I} \widehat{I} + a_{i 2} (\widehat{I} \widehat{X} + (\widehat{\sigma}^{\dag} \widehat{\sigma} + \mathit{h.c.}) )) \\
& + a_{j 1 2} \widehat{X} \widehat{I} \widehat{I}
\end{aligned}
\end{equation*}
\item three dimension:
\begin{equation*}
\begin{aligned}
\underline{\underline{A}} & \Leftrightarrow
\widehat{m} \widehat{m} (a_{1} \widehat{I} \widehat{I} + a_{i 1} (\widehat{I} \widehat{X} + (\widehat{\sigma}^{\dag} \widehat{\sigma} + \mathit{h.c.}) )) \\
& + \widehat{m} \widehat{n} (a_{2} \widehat{I} \widehat{I} + a_{i 2} (\widehat{I} \widehat{X} + (\widehat{\sigma}^{\dag} \widehat{\sigma} + \mathit{h.c.}) )) \\
& + a_{j 1 2} \widehat{m} \widehat{X} \widehat{I} \widehat{I} + a_{j 3 4} \widehat{n} \widehat{X} \widehat{I} \widehat{I} \\
& + \widehat{n} \widehat{m} (a_{3} \widehat{I} \widehat{I} + a_{i 3} (\widehat{I} \widehat{X} + (\widehat{\sigma}^{\dag} \widehat{\sigma} + \mathit{h.c.}) )) \\
& + \widehat{n} \widehat{n} (a_{4} \widehat{I} \widehat{I} + a_{i 4} (\widehat{I} \widehat{X} + (\widehat{\sigma}^{\dag} \widehat{\sigma} + \mathit{h.c.}) )) \\
& + a_{k 1 3} \widehat{X} \widehat{m} \widehat{I} \widehat{I} + a_{k 2 4} \widehat{X} \widehat{n} \widehat{I} \widehat{I}
\end{aligned}
\end{equation*}
\end{itemize}

In the simple \Cref{eq_pde_simple} case: $ a_{1} = a_{2} = a_{3} = a_{4} = -4 $, $ a_{i 1} = a_{i 2} = a_{i 3} = a_{i 4} = a_{j 1 2} = a_{j 3 4} = a_{k13} = a_{k 2 4} = 1 $.
In this basic case, some matrices can be implemented with the same operator:
\begin{equation*}
\begin{aligned}
\underline{\underline{A}} & \Leftrightarrow
\widehat{I} \widehat{I} (a \widehat{I} \widehat{I}  + a_{i} (\widehat{I} \widehat{X} + (\widehat{\sigma}^{\dag} \widehat{\sigma} + \mathit{h.c.}) )) \\
& + a_{j} \widehat{I} \widehat{X} \widehat{I} \widehat{I} + a_{k} \widehat{X} \widehat{I} \widehat{I} \widehat{I}
\end{aligned}
\end{equation*}

By induction, the number of nodes per line can be implemented thanks to a logarithmic number of $ \{ (\widehat{\sigma}^{\dag} \widehat{\sigma} \widehat{\sigma} + \mathit{h.c.}); (\widehat{\sigma}^{\dag} \widehat{\sigma} \widehat{\sigma} \widehat{\sigma} + \mathit{h.c.}); (\widehat{\sigma}^{\dag} \widehat{\sigma} \widehat{\sigma} \widehat{\sigma} \widehat{\sigma} + \mathit{h.c.}) ... \} $ gates with respect to the matrice size.
Since each new gate needs one control more than the previous one, the two-qubit gate for each new gate increased linearly with the qubit number and so in a logarithmic fashion with respect to the matrix size.
It leads to a two-qubit gate number:
\begin{equation}
    nb_{2qb} \propto \sum_{i = 1}^{n = log_{2}(N)} i = \frac{\log_{2}(N)^{2} + \log_{2}(N)}{2}
\end{equation}
scaling with $N$ the matrix size.
\cite{ty_double-logarithmic_2024} propose a more efficient (double-logarithmic in depth) strategy to \ac{be} the \Cref{eq_pde_simple} in one dimention.
The advantage of the method proposed in this article is its strategy simplicity, addressed matrix versatility, and ability to also produce Hamiltonian simulations.

Recent work \cite{sato_hamiltonian_2024} proposed a specific case of the Hamiltonian simulation structure to model \ac{pde}.
They specify the choice of doing Hamiltonian simulation instead of \ac{be} and can not address the matrix component individually.
Their work is the latest improvement, which explicit the
gate construction, of the complex Schrodingerisation concept \cite{jin_quantum_2022}.

\subsubsection{Boundary Condition}
While the Dirichlet boundary condition, used when the function value is known at some points, is determined by the value of the second member.
For Neuman boundary conditions, the solution is constrained by his derivatives, for instance:
\begin{equation}
    \frac{\partial}{\partial x} f_{i} = \gamma \Leftrightarrow f_{i \pm 1} = f_{i \mp 1} \pm 2 d \gamma
\end{equation}
To deal with it, the formalism brought by the present paper allows specific components to be addressed for only one extra exponential Hermitian gate.
It is done for $i = 3$ in the first node line and for the whole second node line in the example matrix $B$.
Similarly, as for the first neighbor, the generalization to lines of terms that interact with their second neighbor is written as: $ \{ \widehat{X} \widehat{I}; (\widehat{\sigma}^{\dag} \widehat{\sigma} + \mathit{h.c.}) \widehat{I}; (\widehat{\sigma}^{\dag} \widehat{\sigma} \widehat{\sigma} + \mathit{h.c.}) \widehat{I}; (\widehat{\sigma}^{\dag} \widehat{\sigma} \widehat{\sigma} \widehat{\sigma} + \mathit{h.c.}) \widehat{I} ... \} $
It can also be beneficial to impose specific conditions such as periodic conditions, as in \cite{ty_double-logarithmic_2024}, for which one needs to change the component controlled by:
\begin{itemize}
\item $ \widehat{m} \widehat{m} \dots \widehat{m} $
\item $ \widehat{n} \widehat{n} \dots \widehat{n} $
\item $ \widehat{\sigma} \widehat{\sigma} \dots \widehat{\sigma} + \mathit{h.c.} $
\end{itemize}
in one dimension or node-line.
Similarly to adding extra node lines, the line on which these conditions are applied can be controlled thanks to an extra tensorial product with $\widehat{m}$ and $\widehat{n}$.
For instance, for the two node-lines case:
\begin{equation*}
    \begin{aligned}
\underline{\underline{B}} & = \begin{bmatrix}
b_{11} & 0 & 0 & b_{i1} & 0 & 0 & 0 & b_{j12} \\
0 & 0 & 0 & b_{124} & 0 & 0 & 0 & 0 \\
0 & 0 & 0 & 0 & 0 & 0 & 0 & 0 \\
b_{i1} & b_{124} & 0 & b_{12} & 0 & 0 & 0 & 0 \\
0 & 0 & 0 & 0 & b_{21} & 0 & b_{ii} & b_{i2} \\
0 & 0 & 0 & 0 & 0 & 0 & 0 & b_{ii} \\
0 & 0 & 0 & 0 & b_{ii} & 0 & 0 & 0 \\
b_{j12} & 0 & 0 & 0 & b_{i2} & b_{ii} & 0 & b_{22}
\end{bmatrix} \\
& \Leftrightarrow b_{11} \widehat{m} \widehat{m} \widehat{m} + b_{12} \widehat{m} \widehat{n} \widehat{n} + b_{21} \widehat{n} \widehat{m} \widehat{m} + b_{22} \widehat{n} \widehat{n} \widehat{n} \\
& + b_{i1} (\widehat{m} \widehat{\sigma} \widehat{\sigma} + \mathit{h.c.}) + b_{i2} (\widehat{n} \widehat{\sigma} \widehat{\sigma} + \mathit{h.c.}) \\
& + b_{j12} (\widehat{\sigma} \widehat{\sigma} \widehat{\sigma} + \mathit{h.c.} ) + b_{124} \widehat{m} \widehat{X} \widehat{n} + b_{ii} \widehat{n} \widehat{X} \widehat{I}
\end{aligned}
\end{equation*}

Another helpful option from this article's formalism is allowing the equation coefficients to be inhomogeneous in space.
Thanks to the formalism proposed in the paper, it is possible with only one extra gate to control the coefficients associated with each line of the modeled system, for instance, to model a difference between two mediums, corresponding to $\alpha$ a spatial function or a different differential equation in different place of the space.

\subsection{Arbitrary Sparse Matrix Decomposition}
The abitily to implement and evaluate states on arbitrary sparse matrices is something of interest for some technics.
This section describes how the operator that allow to implement an arbitrary component amplitude of an Hermitian can be constructed:
$$ \widehat{H_{p}} = \sum_{a, b} w_{a, b} (\ket{\mathrm{bin}[a]}\bra{\mathrm{bin}[b]} + \mathit{h.c.}) $$
It is only required to construct the single component transition from the \acl{scb} operators: 
\begin{equation*}
\begin{aligned}
    & (\ket{\mathrm{bin}[a]}\bra{\mathrm{bin}[b]} + \mathit{h.c.}) = (\bigotimes_{i} \widehat{n}_{i} \bigotimes_{j} \widehat{m}_{j} \bigotimes_{k} \widehat{\sigma}_{k} \bigotimes_{l} \widehat{\sigma}_{l}^{\dag} + \mathit{h.c.}) \\
    & \text{with } i \neq j \neq k \neq l
\end{aligned}
\end{equation*}

It can be done simply using the table \Cref{table_single_component} as illustrated with the exemple with $ a = 1222 $ and $ b = 1145 $ by:
\begin{equation*}
    \begin{aligned}
        & \ket{\mathrm{bin}[1222]}\bra{\mathrm{bin}[1145]} \\
        = & \ket{10011000110}\bra{10001111001} \\
        = & \widehat{n} \widehat{m} \widehat{m} \widehat{\sigma}^{\dag} \widehat{n} \widehat{\sigma} \widehat{\sigma} \widehat{\sigma} \widehat{\sigma}^{\dag} \widehat{\sigma}^{\dag} \widehat{\sigma}
    \end{aligned}
\end{equation*}

\begin{table}[ht]
    \caption{Single component transition form the single component operators}
    \begin{center}
        \begin{tabular}{|c|c|c|c|c|}
            \hline
            \textbf{Index} & \multicolumn{4}{|c|}{\textbf{Digit value}} \\
            \hline
            $\mathrm{bin}[a]$ & $0$ & $1$ & $0$ & $1$ \\
            \hline
            $\mathrm{bin}[b]$ & $0$ & $1$ & $1$ & $0$ \\
            \hline
            \textbf{Operator} & $\widehat{m}$ & $\widehat{n}$ & $\widehat{\sigma}$ & $\widehat{\sigma}^{\dag}$ \\
            \hline
        \end{tabular}
    \end{center}
    \label{table_single_component}
\end{table}

\subsection{Deal with non-Hermitian Matrix as in \acl{qlsp}}
\label{section_qlsp}
To treat non-Hermitian matrix $\widehat{A}$, \cite{harrow_quantum_2009} and \cite{berry_efficient_2007} propose to use the following matrix transformation:
\begin{equation}
    \widehat{H} = \widehat{\sigma}_{0}^{\dag} \widehat{A} + \mathit{h.c.}
\end{equation}
and the vector $\ket{a}$ transformation:
\begin{equation}
    \ket{\psi} = \ket{0} \ket{a}
\end{equation}
so that:
\begin{equation}
    \widehat{H} \ket{\psi} = \ket{1} \widehat{A} \ket{a}
\end{equation}
This example shows how direct formalism avoids multiplying by at least four the number of terms to exponentials (or to sum) in the \ac{be} with Pauli decomposition techniques:
\begin{equation}
    \widehat{H} = \frac{\widehat{X} - i \widehat{Y}}{2} \widehat{A} + \frac{\widehat{X} + i \widehat{Y}}{2} \widehat{A}^{\dag}
\end{equation}

\section{Discussion}
\subsection{Identifying the Most Efficient Strategy}
On \ac{nisq} devices, in order to know whether the usual or direct Hamiltonian simulation strategy leads to the shallower circuit, a suitable method is to express the depth of the Hamiltonian simulation of the two strategies transpiled to the native \ac{qpu} set of gates.

\subsection{Compatibility with Trotterization and Variational Algorithms Variants}
Most Trotterization and Variational Algorithms Variants can be adapted to direct Hamiltonian simulation.
This is natively true for variational algorithms that seek the optimal order of the product formula, such as \cite{ikeda_measuring_2024}, or the optimal ordering of the operators \cite{grimsley_adaptive_2019}.
Algorithms that mitigate the Trotter error, such as \ac{mpf} \cite{childs_hamiltonian_2012,low_well-conditioned_2019, aftab_multi-product_2024} can also be used with the direct method.

Other techniques whose constructions are based on product formulas such as corrected product formulas \cite{bagherimehrab_faster_2024}, qDRIFT \cite{campbell_random_2019} or imaginary time evolution \cite{leadbeater_non-unitary_2023} might be adapted to direct Hamiltonian simulation.

Finally, with a good optimizer and the good terms gathered, the Hamiltonian compilation \cite{mukhopadhyay_synthesizing_2023} by multiple Pauli-strings diagonalization \cite{berg_circuit_2020} proposal bridges the gap between the usual and direct strategies.
Using a similar strategy to gather more terms could be very interesting but would make an algorithm harder to automate.

\section{Conclusion}
This paper presents a different strategy called direct Hamiltonian simulation to construct quantum algorithms primitives.
It comes with a simple formalism that directly allows the construction of the Hamiltonian simulation of the tensorial product of the \acl{scb} plus  Pauli-strings.
The number of unitary rotations needed per Trotter-step equals the number of terms in the initial formalism instead of having a number of unitary matrices growing exponentially with the number of mapped terms.
It leads to a quantum circuit with a competitive depth, number of two-qubit gates, and number of arbitrary rotations with respect to the usual method.
It can also be seen as a unification of the Hamiltonian simulation techniques developed in different fields, allowing the recovery of state-of-the-art results for at least the three presented domains.
The improvement comes in the simplicity and modularity of the method.

For instance, this technique allows (under certain assumptions) the exponential reduction, with respect to the \ac{hubo} order, of a problem Hamiltonian simulation quantum circuit depth without any accuracy loss.
Chemical simulation can also be improved thanks to the operator's easier interpretation, leading to \ac{uccsd} ansatz interpretable as a series of electronic transitions.
This formalism can address finite-difference problems on regular grids in a very versatile way with a tractable number of gates.
It is well adapted to transform the non-Hermitian matrix without changing the number of terms, for instance, in order to treat \ac{qlsp} problems.

It can also be used to construct \ac{be} of the operator of interest with at most six unitary matrices per term.

The authors hope the formalism proposed in this article will simplify the implementing of existing and new problems into quantum computers.

\bibliographystyle{ieeetr}
\bibliography{biblio}


\newpage

\section*{\textbf{ \Large Appendix}}

\begin{abstract}
    The basic appendix shows calculation details and quantum circuit construction by direct applications, allowing notation clarifications. \\
    The annex is more foccussed on the formalism than the initial paper; it aims to show the natural consequences of the associated paper such as the measurement of expectation values with fewer observables.
    It also provides an alternative interpretation of the main paper's technique for basis transformation, which seamlessly extends to an arbitrary one-qubit gate acting on two selected states.
\end{abstract}

\section{Appendix: Calculation Details and Quantum Circuits}
\label{annex}
\subsection{Calculation Details}
\subsubsection{\acl{hubo}}
The link between control-phase gate and the exponential of the number of exitation operator:

$ \widehat{m} = \ket{0} \bra{0} =
\begin{bmatrix}
    1 & 0 \\
    0 & 0
\end{bmatrix} = \widehat{\sigma} \widehat{\sigma}^{\dag} $;

$ \widehat{n} = \ket{1} \bra{1} =
\begin{bmatrix}
    0 & 0 \\
    0 & 1
\end{bmatrix} = \widehat{\sigma}^{\dag} \widehat{\sigma} $;

$ e^{i t \widehat{n}} =
\begin{bmatrix}
    1 & 0 \\
    0 & e^{i t}
\end{bmatrix} = \widehat{P}(t) $ ;

$ \widehat{n} \widehat{n} = \ket{1 1} \bra{1 1} =
\begin{bmatrix}
    0 & 0 & 0 & 0 \\
    0 & 0 & 0 & 0 \\
    0 & 0 & 0 & 0 \\
    0 & 0 & 0 & 1
\end{bmatrix} $ ;

$ e^{i t \widehat{n} \widehat{n}} =
\begin{bmatrix}
    1 & 0 & 0 & 0 \\
    0 & 1 & 0 & 0 \\
    0 & 0 & 1 & 0 \\
    0 & 0 & 0 & e^{i t}
\end{bmatrix} = \widehat{CP}(t) $

Developed calculation of the three first-order excitation numbers mapped to a \ac{lcu} notation switches:

\begin{equation*}
    \widehat{Z} = \widehat{I} - 2 \widehat{n}
\end{equation*}

\begin{equation*}
    \begin{aligned}
        \widehat{Z Z} & = (\widehat{I} - 2 \widehat{n}_{i}) (\widehat{I} - 2 \widehat{n}_{j}) \\
        & = \widehat{I} + 4 \widehat{n} \widehat{n} - 2 ( \widehat{n}_{i} + \widehat{n}_{j} )
    \end{aligned}
\end{equation*}

\begin{equation*}
    \begin{aligned}
        \widehat{Z} \widehat{Z} \widehat{Z} = & (\widehat{I} + 4 \widehat{n}_{i} \widehat{n}_{j} - 2 ( \widehat{n}_{i} + \widehat{n}_{j} ) ) (\widehat{I} - 2 \widehat{n}_{k})  \\
        = & \widehat{I} + 4 \widehat{n}_{i} \widehat{n}_{j} - 2 ( \widehat{n}_{i} + \widehat{n}_{j} ) - 2 \widehat{n}_{k} - 8 \widehat{n} \widehat{n} \widehat{n} + 4 ( \widehat{n}_{i} \widehat{n}_{k} + \widehat{n}_{j} \widehat{n}_{k} ) \\
        = & \widehat{I} - 8 \widehat{n} \widehat{n} \widehat{n} + 4 ( \widehat{n}_{i} \widehat{n}_{j} + \widehat{n}_{i} \widehat{n}_{k} + \widehat{n}_{j} \widehat{n}_{k} ) - 2 ( \widehat{n}_{i} + \widehat{n}_{j} + \widehat{n}_{k} )
    \end{aligned}
\end{equation*}

\begin{equation*}
    \begin{aligned}
        \widehat{n} \widehat{n} & = ( \frac{\widehat{I} - \widehat{Z}_{i}}{2} ) ( \frac{\widehat{I} - \widehat{Z}_{j}}{2} ) \\
        & = \frac{1}{4} ( \widehat{I} + \widehat{Z Z} - \widehat{Z}_{i} - \widehat{Z}_{j} )
    \end{aligned}
\end{equation*}

\begin{equation*}
    \begin{aligned}
        \widehat{n} \widehat{n} \widehat{n} = & \frac{1}{8} ( \widehat{I} + \widehat{Z Z}_{i j} - \widehat{Z}_{i} - \widehat{Z}_{j} ) ( \widehat{I} - \widehat{Z}_{k} ) \\
        = & \frac{1}{8} ( \widehat{I} + \widehat{Z Z}_{i j} - \widehat{Z}_{i} - \widehat{Z}_{j}
        - \widehat{Z}_{k} - \widehat{Z Z Z} \\
        & + \widehat{Z Z}_{i k} + \widehat{Z Z}_{j k} ) \\
        = & \frac{1}{8} ( \widehat{I} - \widehat{Z Z Z} + \widehat{Z Z}_{i j} + \widehat{Z Z}_{i k} + \widehat{Z Z}_{j k} \\
        & - ( \widehat{Z}_{i} + \widehat{Z}_{j} + \widehat{Z}_{k} ) )
    \end{aligned}
\end{equation*}

\begin{table*}[ht]
\caption{Three first order \ac{hubo} Hamiltonian simulation}
\begin{center}
\resizebox{\linewidth}{!}{
\includegraphics{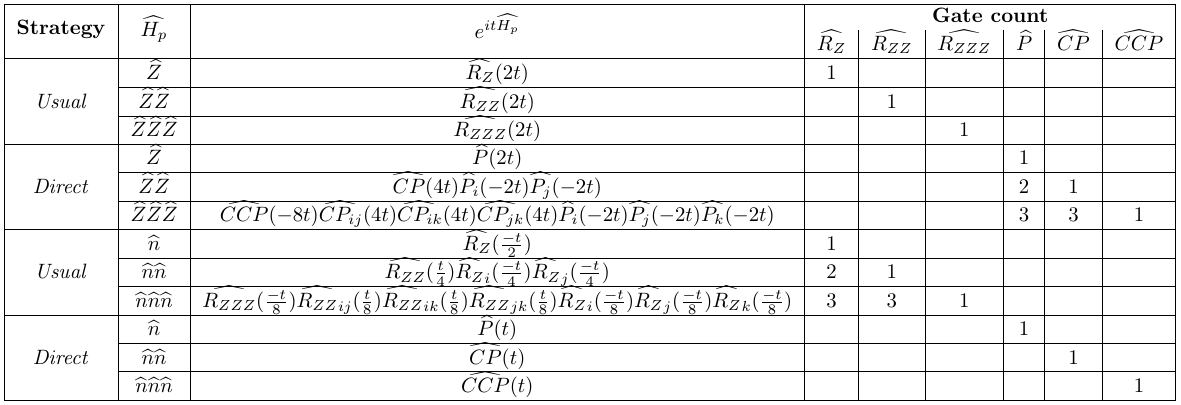}}
\end{center}
\label{table_hubo}
\end{table*}

\subsubsection{Fermionic Hamiltonian}
In the following section, $i \neq j$ and $k \neq l$.
Different equivalent notations of ladder operator composition:

$ \widehat{\sigma} = \ket{0} \bra{1} =
\begin{bmatrix}
    0 & 0 \\
    1 & 0
\end{bmatrix} $ ;

$ \widehat{A}_{1} = \widehat{\sigma}_{i}^{\dag} \widehat{\sigma}_{j} + \mathit{h.c.} = \ket{10} \bra{01} + \mathit{h.c.} =
\begin{bmatrix}
    0 & 0 & 0 & 0 \\
    0 & 0 & 1 & 0 \\
    0 & 1 & 0 & 0 \\
    0 & 0 & 0 & 0 
\end{bmatrix} $ ;

$\widehat{B} = \alpha \widehat{\sigma}^{\dag} \widehat{\sigma} + \beta \widehat{\sigma}^{\dag} \widehat{\sigma}^{\dag} + \mathit{h.c.} = 
\begin{bmatrix}
    0 & 0 & 0 & \beta \\
    0 & 0 & \alpha & 0 \\
    0 & \alpha & 0 & 0 \\
    \beta & 0 & 0 & 0
\end{bmatrix} $ ;

\begin{equation*}
\widehat{H}_{1} = \widehat{a}_{i}^{\dag} \widehat{a}_{j} + \mathit{h.c.} =
\begin{bmatrix}
    \widehat{A}_{1} & 0 \\
    0 & - \widehat{A}_{1}
\end{bmatrix}
\begin{matrix}
    \ket{\psi}_{i j} \ket{\mathrm{Par}[\psi] \text{ is odd}}_{i + 1 \rightarrow j -1} \\
    \ket{\psi}_{i j} \ket{\mathrm{Par}[\psi] \text{ is even}}_{i + 1 \rightarrow j -1}
\end{matrix}
\end{equation*}

With $\ket{\mathrm{Par}[\psi] \text{ is odd}}_{i + 1 \rightarrow j -1}$ (similarly for $\ket{\mathrm{Par}[\psi] \text{ is even}}_{i + 1 \rightarrow j -1}$) denote all the states with an odd number of $1$ in the computational basis between $i$ and $j$ qubits.

$ e^{ i t \widehat{A}_{1}} =
\begin{bmatrix}
    1 & 0 & 0 & 0 \\
    0 & \cos(t) & i \sin(t) & 0 \\
    0 & i \sin(t) & \cos(t) & 0 \\
    0 & 0 & 0 & 1 
\end{bmatrix} $ ;

$ \widehat{\sigma}_{i}^{\dag} \widehat{\sigma}_{j}^{\dag} \widehat{\sigma}_{k} \widehat{\sigma}_{l} + \mathit{h.c.} = \ket{1100} \bra{0011} + \mathit{h.c.} $ ;

\begin{equation*}
\begin{aligned}
e^{ i t (\widehat{\sigma}^{\dag} \widehat{\sigma}^{\dag} \widehat{\sigma} \widehat{\sigma} + \mathit{h.c.})} = &
\cos(t) (\ket{0011} \bra{0011} + \ket{1100} \bra{1100}) \\
& + i \sin(t) (\ket{0011} \bra{1100} + \mathit{h.c.}) \\
& + \sum_{\substack{j = 0\\ \text{s.t. } \mathrm{bin}[j] \neq \{ 0011 ; 1100 \}}} ^{15} \ket{\mathrm{bin}[j]} \bra{\mathrm{bin}[j]}
\end{aligned}
\end{equation*}

Using $ e^{- i \theta \widehat{n}_{i} \widehat{H} } = \widehat{C e^{- i \theta H}} $ :

$ \widehat{CR_{X}}(- 2 \theta) = e^{- i \theta \widehat{n}_{i} ( \widehat{\sigma}_{j} + \mathit{h.c.} ) } $ ;

$ \widehat{Ce^{ i t A_{1}}} = e^{- i t \widehat{n}_{i} ( \widehat{\sigma}_{j}^{\dag} \widehat{\sigma}_{k} + \mathit{h.c.} ) } $ ;

\subsection{Quantum Circuit}

\subsubsection{Pauli-String Hamiltonian Simulation}
Efficient quantum circuit construction of Pauli-string Hamiltonian simulation thanks to the parity report on one qubit.

\begin{figure}[H]
\begin{center}
\resizebox{\linewidth}{!}{
\includegraphics{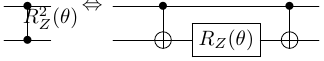}}
\end{center}
\caption{$\widehat{R_{ZZ}}$ gate efficient decomposition.}
\end{figure}

\begin{figure}[H]
\begin{center}
\resizebox{\linewidth}{!}{\includegraphics{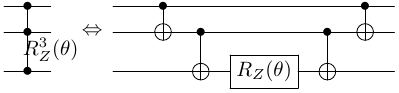}}
\end{center}
\caption{$\widehat{R_{ZZZ}}$ gate efficient decomposition.}
\end{figure}

\begin{figure}[H]
\begin{center}
\resizebox{\linewidth}{!}{\includegraphics{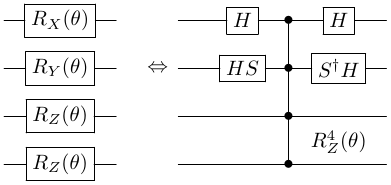}}
\end{center}
\caption{$\widehat{R_{XYZZ}}$ gate efficient decomposition.}
\end{figure}

\subsubsection{Fermionic Operator from the Qubit Operator}.

\begin{figure}[H]
\begin{center}
\resizebox{9 cm}{!}{\includegraphics{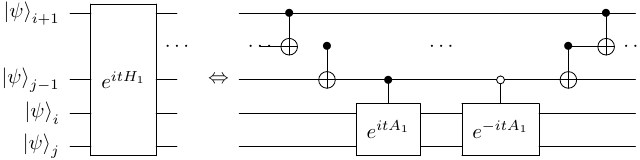}}
\end{center}
\caption{$ e^{i t \widehat{H}_{1}} $ decomposition with $ e^{i t \widehat{A}_{1}} $, $i < j$.}
\label{ligne_la_plus_stylee}
\end{figure}

\begin{figure}[H]
\begin{center}
\resizebox{9 cm}{!}{\includegraphics{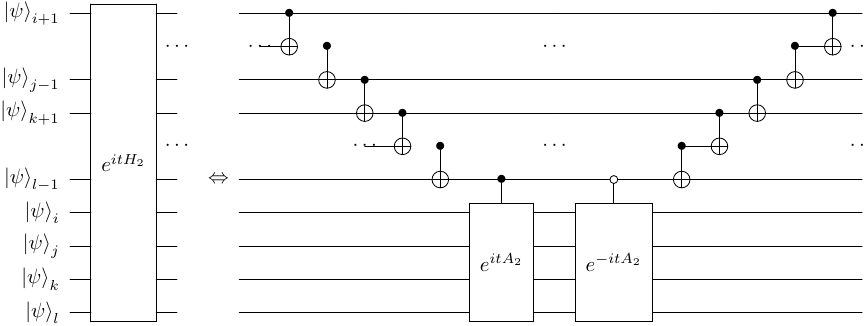}}
\end{center}
\caption{$ e^{i t \widehat{H}_{2}} $ decomposition with $ e^{i t \widehat{A}_{2}} $, $i < j < k < l$.}
\label{ligne_la_plus_stylee_bis}
\end{figure}

\subsubsection{Constructing the Gates that act in-Between Qubits}
In order to represent a gate acting in-between qubits, the following convention is used:
$$ \widehat{C^{n} U} \{\ket{\psi_{1}}; \ket{\psi_{2}}\} = 
\begin{bmatrix}
    1 & 0 & 0 \\
    0 & U_{11} & U_{12} \\
    0 & U_{21} & U_{22}
\end{bmatrix}
\begin{matrix}
    \ket{\psi_{\perp}} \\
    \ket{\psi_{1}} \\
    \ket{\psi_{2}}
\end{matrix} $$ mean that:
\begin{itemize}
    \item the gate acts on $n+1$ qubits.
    \item all the states except the one in the curly bracket, denoted by $\ket{\psi_{\perp}}$.
    \item the states in the curly brackets are affected by gate $\widehat{U}$.
\end{itemize}
The following list shows exemples of gates acting in-between qubits:
\begin{itemize}
    \item $ \widehat{PP}\{ \ket{01}; \ket{10} \} = 
    \begin{bmatrix}
        1 & 0 & 0 & 0\\
        0 & e^{i \theta} & 0 & 0 \\
        0 & 0 & e^{i \theta} & 0 \\
        0 & 0 & 0 & 1
    \end{bmatrix}
    \begin{matrix}
        \ket{00} \\
        \ket{01} \\
        \ket{10} \\
        \ket{11}
    \end{matrix} $
    \item $ \widehat{CR_{Z}}\{ \ket{01}; \ket{10} \} = 
    \begin{bmatrix}
        1 & 0 & 0 & 0\\
        0 & e^{- i \theta /2} & 0 & 0 \\
        0 & 0 & e^{i \theta /2} & 0 \\
        0 & 0 & 0 & 1
    \end{bmatrix}
    \begin{matrix}
        \ket{00} \\
        \ket{01} \\
        \ket{10} \\
        \ket{11}
    \end{matrix} $
    \item $
    \begin{matrix}
        \\
        \widehat{CR_{X}}\{ \ket{01}; \ket{10} \} = \\
        \\
        \\
        \\
        = e^{- i \frac{t}{2} \widehat{A_{1}}}
    \end{matrix} 
    \begin{bmatrix}
        1 & 0 & 0 & 0\\
        0 & \cos & - i \sin & 0 \\
        0 & - i \sin & \cos & 0 \\
        0 & 0 & 0 & 1
    \end{bmatrix}
    \begin{matrix}
        \ket{00} \\
        \ket{01} \\
        \ket{10} \\
        \ket{11}
    \end{matrix} $
    \item $ \widehat{CR_{Y}}\{ \ket{01}; \ket{10} \} = 
    \begin{bmatrix}
        1 & 0 & 0 & 0\\
        0 & \cos & - \sin & 0 \\
        0 & \sin & \cos & 0 \\
        0 & 0 & 0 & 1
    \end{bmatrix}
    \begin{matrix}
        \ket{00} \\
        \ket{01} \\
        \ket{10} \\
        \ket{11}
    \end{matrix} $
    \item $
    \begin{matrix}
        \\
        \widehat{C^{3}R_{X}}\{ \ket{0011}; \ket{1100} \} = \\
        \\
        \\
        = e^{- i \frac{t}{2} (\widehat{\sigma}^{\dag} \widehat{\sigma}^{\dag} \widehat{\sigma} \widehat{\sigma} + \mathit{h.c.})} \\
        = e^{- i \frac{t}{2} \widehat{A}_{2}}
    \end{matrix}
    \begin{bmatrix}
        1 & 0 & 0 \\
        0 & \cos & - i \sin \\
        0 & - i \sin & \cos
    \end{bmatrix}
    \begin{matrix}
        \ket{\psi_{\perp}} \\
        \ket{0011} \\
        \ket{1100}
    \end{matrix} $
    \item $
    \begin{matrix}
        \\
        \widehat{CR_{X}}\{ \ket{00}; \ket{11} \} = \\
        \\
        \\
        \\
        = e^{- i \frac{t}{2} (\widehat{\sigma}^{\dag} \widehat{\sigma}^{\dag} + \mathit{h.c.})}
    \end{matrix} 
    \begin{bmatrix}
        \cos & 0 & 0 & - i \sin \\
        0 & 1 & 0 & 0 \\
        0 & 0 & 1 & 0 \\
        - i \sin & 0 & 0 & \cos
    \end{bmatrix}
    \begin{matrix}
        \ket{00} \\
        \ket{01} \\
        \ket{10} \\
        \ket{11}
    \end{matrix} $
    \item $
    \begin{matrix}
        \\
        \widehat{CR_{X}}\{ \ket{00}; \ket{11} \} = \\
        \\
        \\
        \\
        = e^{- i \frac{t}{2} (\widehat{\sigma}^{\dag} \widehat{\sigma}^{\dag} + \mathit{h.c.})}
    \end{matrix} 
    \begin{bmatrix}
        \cos & 0 & 0 & - i \sin \\
        0 & 1 & 0 & 0 \\
        0 & 0 & 1 & 0 \\
        - i \sin & 0 & 0 & \cos
    \end{bmatrix}
    \begin{matrix}
        \ket{00} \\
        \ket{01} \\
        \ket{10} \\
        \ket{11}
    \end{matrix} $
    \item $
    \begin{matrix}
        e^{- i \widehat{B}} = \\
        \\
        \\
        \\
    \end{matrix} 
    \begin{bmatrix}
        \cos(\beta) & 0 & 0 & - i \sin(\beta) \\
        0 & \cos(\alpha) & - i \sin(\alpha) & 0 \\
        0 & - i \sin(\alpha) & \cos(\alpha) & 0 \\
        - i \sin(\beta) & 0 & 0 & \cos(\beta)
    \end{bmatrix}
    \begin{matrix}
        \ket{00} \\
        \ket{01} \\
        \ket{10} \\
        \ket{11}
    \end{matrix} $
\end{itemize}
$e^{- i t \widehat{\sigma}^{\dag} \widehat{\sigma}^{\dag} + \mathit{h.c.}}$ is relevant to model strongly correlated electron models \cite[Fig.10]{wecker_solving_2015}, that is why the associated quantum circuit is also given \Cref{bonus}.

\begin{figure}[H]
\begin{center}
\resizebox{\linewidth}{!}{\includegraphics{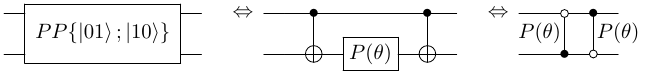}}
\end{center}
\caption{$\widehat{PP}\{ \ket{01}; \ket{10} \}$ gate decomposition.}
\end{figure}

\begin{figure}[H]
\begin{center}
\resizebox{\linewidth}{!}{\includegraphics{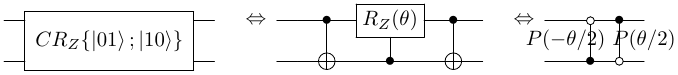}}
\end{center}
\caption{$\widehat{CR_{Z}}\{ \ket{01}; \ket{10} \}$ gate decomposition.}
\label{mag12}
\end{figure}

\begin{figure}[H]
\begin{center}
\resizebox{\linewidth}{!}{\includegraphics{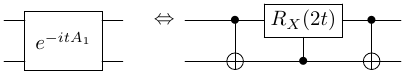}}
\end{center}
\caption{$e^{- i t \widehat{A_{1}}}$ gate decomposition.}
\label{mag22}
\end{figure}

\begin{figure}[H]
\begin{center}
\resizebox{\linewidth}{!}{\includegraphics{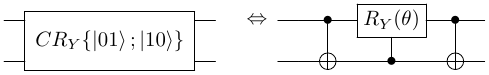}}
\end{center}
\caption{$\widehat{CR_{Y}}\{ \ket{01}; \ket{10} \}$ gate decomposition.}
\label{mag32}
\end{figure}

\begin{figure}[H]
\begin{center}
\resizebox{\linewidth}{!}{\includegraphics{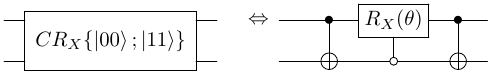}}
\end{center}
\caption{$\widehat{CR_{X}}\{ \ket{00}; \ket{11} \}$ gate decomposition.}
\label{bonus}
\end{figure}

\begin{figure}[H]
\begin{center}
\resizebox{\linewidth}{!}{\includegraphics{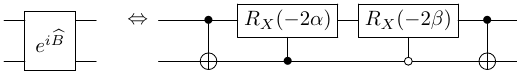}}
\end{center}
\caption{$e^{i \widehat{B}}$ gate decomposition.}
\label{bonus2}
\end{figure}

\begin{figure}[H]
\begin{center}
\resizebox{\linewidth}{!}{\includegraphics{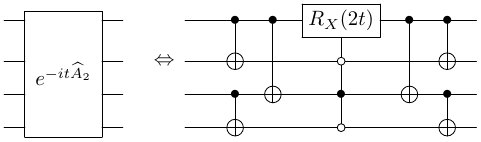}}
\end{center}
\caption{$e^{- i t \widehat{A}_{2}}$ gate decomposition.}
\label{mag42}
\end{figure}

\subsubsection{Controlled Gates that act in-Between Qubits}
\begin{figure}[H]
\begin{center}
\resizebox{\linewidth}{!}{\includegraphics{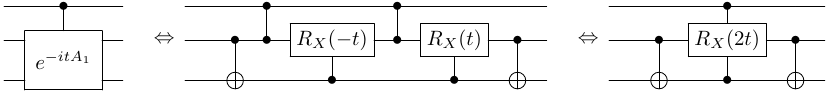}}
\end{center}
\caption{$\widehat{Ce^{- i t A_{1}}}$ gate decomposition, the second circuit take advantage of: $ \widehat{C_{a} R_{X/Y}}(\theta) = \widehat{R_{X/Y}}(\theta) \widehat{C_{a}Z} \widehat{R_{X/Y}}(- \theta) \widehat{C_{a}Z} $}
\label{ctrl_gate}
\end{figure}

\begin{figure}[H]
\begin{center}
\resizebox{\linewidth}{!}{\includegraphics{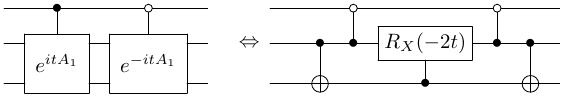}}
\end{center}
\caption{$e^{\pm i t \widehat{A_{1}}}$ gate decomposition.}
\label{ctrl_sign2}
\end{figure}

The previous gates are relatively easy to control \Cref{ctrl_gate}.
Since the goal is only to control the rotation sign: $$ e^{\pm i t \widehat{A_{1}}} = \widehat{Ce^{i t A_{1}}} \widehat{X} \widehat{Ce^{- i t A_{1}}} \widehat{X} $$, a simpler solution is illustrated in the following circuit, it can be easily extended to the other in-between qubit gates \Cref{ctrl_sign2}.

\begin{figure}[H]
\begin{center}
\resizebox{\linewidth}{!}{\includegraphics{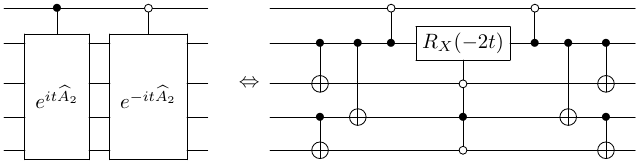}}
\end{center}
\caption{$e^{\pm i t \widehat{A}_{2}}$ gate decomposition.}
\label{ctrl_sign_big}
\end{figure}

\begin{figure}[H]
\begin{center}
\resizebox{7 cm}{!}{\includegraphics{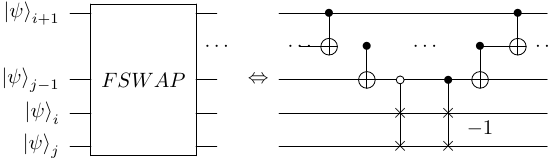}}
\end{center}
\caption{$ \widehat{FSWAP} $ decomposition with $\widehat{sSWAP}$.}
\label{fswap}
\end{figure}

\begin{figure}[H]
\begin{center}
\resizebox{\linewidth}{!}{\includegraphics{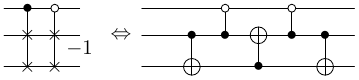}}
\end{center}
\caption{Controll-sign-SWAP gate decomposition.}
\label{sswap}
\end{figure}

The FSWAP gate is defined as \cite{kivlichan_quantum_2018}: $$ \widehat{FSWAP}_{ij} = \widehat{I} - \widehat{a}^{\dag}_{i} \widehat{a}_{i} - \widehat{a}^{\dag}_{j} \widehat{a}_{j} + \widehat{a}^{\dag} \widehat{a} + \mathit{h.c.} $$
\Cref{fswap} and \Cref{sswap} illustrate a naïve construction of this gate.

\section{Annex}

\subsection{Pauli-string Deepness Optimization}
Two equivalent changes of basis were proposed for the \ac{scb} operator Hamiltonian Simulation: one straightforward and one with the optimized gate deepness.
Similarly, the deepness of the parity report of the Pauli-string can be optimized.
As only the parity report on the qubit on which the $Z$ (or $RZ$) gate is applied matters, a similar pyramidal structure can be used to report the parity after the change of basis.

\begin{figure}[H]
    \begin{center}
        \resizebox{\linewidth}{!}{\includegraphics{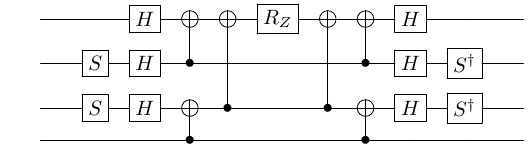}}
    \end{center}
    \caption{$ e^{i \theta \widehat{X} \widehat{Y} \widehat{Y} \widehat{Z}} $ parity report depth optimization with a pyramidal structure.}
    \label{ps}
\end{figure}

\subsection{Arbitrary gate application to two chosen states of the computational basis}
The most general definition of a gate acting in between qubits is:
\begin{equation*}
    \begin{aligned}
        & \widehat{C^{n} V} \{\ket{\mathrm{bin}[a]}; \ket{\mathrm{bin}[b]}\} \\
        = &
        \begin{bmatrix}
            1 & 0 & 0 \\
            0 & V_{11} & V_{12} \\
            0 & V_{21} & V_{22}
        \end{bmatrix}
        \begin{matrix}
            \ket{\psi_{\perp}} \\
            \ket{\mathrm{bin}[a]} \\
            \ket{\mathrm{bin}[b]}
        \end{matrix} \\
        = & \widehat{U_{n \sigma}}^{\dag} \widehat{C^{n} V} \widehat{U_{n \sigma}}
    \end{aligned}
\end{equation*}

\begin{figure}[tb]
    \begin{center}
        \resizebox{\linewidth}{!}{\includegraphics{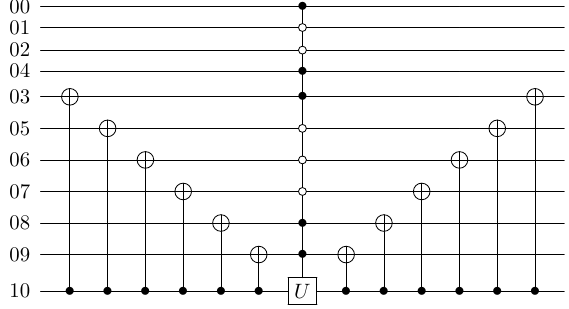}}
    \end{center}
    \caption{$ \widehat{C^{n} U} \{\ket{\mathrm{bin}[1222]}; \ket{\mathrm{bin}[1145]}\} $  gate decomposition, the number on the left of the qubit contains the qubit index.}
    \label{arbi_exemple}
\end{figure}

Interestingly, such a gate can be constructed if the unitary $\widehat{V}$ is known.
As the change of basis that allows access to $ \{\ket{\mathrm{bin}[a]}; \ket{\mathrm{bin}[b]}\} $ is known, it is only needed to apply this gate in the state of interest.
\Cref{arbi_exemple} illustrates it with $ a = 1222 $ and $ b = 1145 $.
It can be used as the beginning of an alternative approach to explain the technics used in this article.

\subsection{Measuring Expectation Value with fewer Observables}
The knowledge of matrix decomposition has two main uses in quantum computing.
It is needed to construct the query that implements the problem and to construct the observables utilized to probe the final quantum state.
Interestingly, with the simplest techniques to evaluate a state, based on many measurements of the state vector, there is no need for the $n$-controlle-gate to make the state evaluation.
The contribution of each shot can thus be computed using:
\begin{equation*}
    \begin{aligned}
        & \bra{\psi} \widehat{H_{P}} \ket{\psi} = \\
        & \quad \sum_{i} \gamma_{i}
            \bra{\psi_{PS}} \widehat{PS}_{i} \ket{\psi_{PS}}
            \bra{\psi_{n \sigma}} (\ket{\mathrm{bin}[a_{i}]}\bra{\mathrm{bin}[b_{i}]} + \mathit{h.c.}) \ket{\psi_{n \sigma}}
    \end{aligned}
\end{equation*}
with $ \bra{\psi_{PS}} \widehat{PS}_{i} \ket{\psi_{PS}} $ evaluated with the usual method,
\begin{equation*}
    \begin{aligned}
        \ket{\psi} & = \ket{\psi_{PS}} \ket{\psi_{n \sigma}}
    \end{aligned}
\end{equation*}
and
\begin{equation*}
    \begin{aligned}
        & \bra{\psi_{n \sigma}} (\ket{\mathrm{bin}[a_{i}]}\bra{\mathrm{bin}[b_{i}]} + \mathit{h.c.}) \ket{\psi_{n \sigma}} = \\
        & \quad \bra{\psi_{n \sigma}} \widehat{U_{n \sigma}}^{\dag} (\ket{\mathrm{bin}[a_{i}]}\bra{\mathrm{bin}[a_{i}]} - \ket{\mathrm{bin}[b_{i}]}\bra{\mathrm{bin}[b_{i}]}) \widehat{U_{n \sigma}} \ket{\psi_{n \sigma}}
    \end{aligned}
\end{equation*}

\begin{figure}[tb]
    \begin{center}
        \resizebox{5cm}{!}{\includegraphics{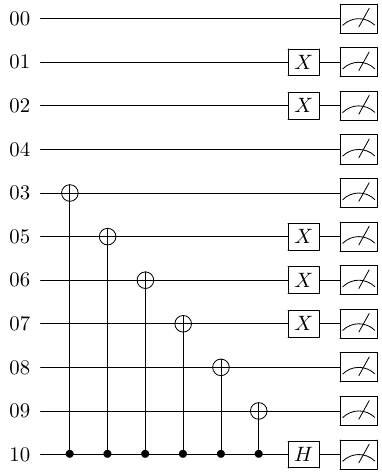}}
    \end{center}
    \caption{$ \ket{\mathrm{bin}[1222]}\bra{\mathrm{bin}[1145]} + \mathit{h.c.} $  change of basis to construct the measurement circuit, the number on the left of the qubit contains the qubit index.}
    \label{basis_change_ex}
\end{figure}

The change of basis $ \widehat{U_{n \sigma}} $ that allows this measurement for the $ a = 1222 $ and $ b = 1145 $ example is illustrated by \Cref{basis_change_ex}.

The naive application of this technique allows the measurement of the expectation value of two body (electrons) energy contributions with $ 2^{4} = 16 $ time fewer observables than the measurement mode of Pauli-strings, using the \acs{nisq} \acs{vqe} algorithm.

\subsection{Tensorial Product Algebra and Commutators}

\begin{table}[ht]
    \caption{Cayley table: tensorial product algebra, each cell contains the $ \widehat{A} \widehat{B} $ matrix with $ \widehat{0} = 
    \begin{bmatrix}
        0 & 0 \\
        0 & 0
    \end{bmatrix} $.}
    \begin{center}
        \begin{tabular}{|c||c|c|c|c|c|c|c|}
            \hline
            $\widehat{A}$ \textbackslash $\widehat{B}$ & $\widehat{m}$ & $\widehat{n}$ & $\widehat{\sigma}$ & $\widehat{\sigma}^{\dag}$ & $\widehat{Z}$ & $\widehat{X}$ & $\widehat{Y}$ \\
            \hline \hline
            $\widehat{m}$ & $\widehat{m}$ & $\widehat{0}$ & $\widehat{0}$ & $\widehat{\sigma}^{\dag}$ & $\widehat{m}$ & $\widehat{\sigma}^{\dag}$ & $-i\widehat{\sigma}^{\dag}$ \\
            \hline
            $\widehat{n}$ & $\widehat{0}$ & $\widehat{n}$ & $\widehat{\sigma}$ & $\widehat{0}$ & $-\widehat{n}$ & $\widehat{\sigma}$ & $i\widehat{\sigma}$ \\
            \hline
            $\widehat{\sigma}$ & $\widehat{\sigma}$ & $\widehat{0}$ & $\widehat{0}$ & $\widehat{n}$ & $\widehat{\sigma}$ & $\widehat{n}$ & $-i\widehat{n}$ \\
            \hline
            $\widehat{\sigma}^{\dag}$ & $\widehat{0}$ & $\widehat{\sigma}^{\dag}$ & $\widehat{m}$ & $\widehat{0}$ & $-\widehat{\sigma}^{\dag}$ & $\widehat{m}$ & $i\widehat{m}$ \\
            \hline
            $\widehat{Z}$ & $\widehat{m}$ & $-\widehat{n}$ & $-\widehat{\sigma}$ & $\widehat{\sigma}^{\dag}$ & $\widehat{I}$ & $-i\widehat{Y}$ & $-i\widehat{X}$ \\
            \hline
            $\widehat{X}$ & $\widehat{\sigma}$ & $\widehat{\sigma}^{\dag}$ & $\widehat{m}$ & $\widehat{n}$ & $i\widehat{Y}$ & $\widehat{I}$ & $i\widehat{Z}$ \\
            \hline
            $\widehat{Y}$ & $i\widehat{\sigma}$ & $-i\widehat{\sigma}^{\dag}$ & $-i\widehat{m}$ & $i\widehat{n}$ & $i\widehat{X}$ & $-i\widehat{Z}$ & $\widehat{I}$ \\
            \hline
        \end{tabular}
    \end{center}
    \label{table_tensorial_product_algebra}
\end{table}

\begin{table}[ht]
    \caption{Commutation relation of Pauli matrices and single-component basis matrices.}
    \begin{center}
        \begin{tabular}{ccc|ccc}
            \multicolumn{6}{c}{\textbf{Commutators}} \\ \\
            $ [\widehat{n}, \widehat{n}] = $ & $ [\widehat{m}, \widehat{m}] = $ & $\widehat{0}$ & $ [\widehat{Z}, \widehat{Z}] = $ & & $\widehat{0}$ \\
            $ [\widehat{\sigma}, \widehat{\sigma}] = $ & $ [\widehat{\sigma}^{\dag}, \widehat{\sigma}^{\dag}] = $ & $\widehat{0}$ & $ [\widehat{X}, \widehat{X}] = $ & $ [\widehat{Y}, \widehat{Y}] = $ & $\widehat{0}$ \\ \\
            $ [\widehat{n}, \widehat{m}] = $ & $ [\widehat{\sigma}, \widehat{\sigma}^{\dag}] = $ & $\widehat{0}$ & $ [\widehat{m}, \widehat{Z}] = $ & $ [\widehat{n}, \widehat{Z}] = $ & $\widehat{0}$ \\ \\
            $ [\widehat{n}, \widehat{\sigma}] = $ & $ [\widehat{\sigma}, \widehat{m}] = $ & $\widehat{\sigma}$ & $ [\widehat{\sigma}, \widehat{Z}] = $ & & $2 \widehat{\sigma}$ \\
            $ [\widehat{m}, \widehat{\sigma}^{\dag}] = $ & $ [\widehat{\sigma}^{\dag}, \widehat{n}] = $ & $\widehat{\sigma}^{\dag}$ & $ [\widehat{Z}, \widehat{\sigma}^{\dag}] = $ & & $2 \widehat{\sigma}^{\dag}$ \\ \\
            $ [\widehat{X}, \widehat{m}] = $ & $ [\widehat{n}, \widehat{X}] = $ & $i \widehat{Y}$ & $ [\widehat{Z}, \widehat{X}] = $ & & $2 i \widehat{Y}$ \\
            $ [\widehat{Y}, \widehat{m}] = $ & $ [\widehat{n}, \widehat{Y}] = $ & $i \widehat{X}$ & $ [\widehat{Y}, \widehat{Z}] = $ & & $2 i \widehat{X}$ \\
            $ [\widehat{\sigma}, \widehat{Y}] = $ & $ [\widehat{\sigma}^{\dag}, \widehat{Y}] = $ & $i \widehat{Z}$ & $ [\widehat{X}, \widehat{Y}] = $ & & $2 i \widehat{Z}$ \\ \\
            $ [\widehat{X}, \widehat{\sigma}] = $ & $ [\widehat{\sigma}^{\dag}, \widehat{X}] = $ & $\widehat{Z}$ & & &  \\ \\
            \hline \\
            \multicolumn{6}{c}{\textbf{Anti-commutators}} \\ \\
            $ \{ \widehat{X}, \widehat{m} \} = $ & $ \{ \widehat{n}, \widehat{X} \} = $ & $\widehat{X}$ & $ \{ \widehat{n}, \widehat{\sigma} \} = $ & $ \{ \widehat{\sigma}, \widehat{m} \} = $ & $\widehat{\sigma}$ \\
            $ \{ \widehat{Y}, \widehat{m} \} = $ & $ \{ \widehat{n}, \widehat{Y} \} = $ & $\widehat{Y}$ & $ \{ \widehat{m}, \widehat{\sigma}^{\dag} \} = $ & $ \{ \widehat{\sigma}^{\dag}, \widehat{n} \} = $ & $\widehat{\sigma}^{\dag}$ \\
            $ \{ \widehat{m}, \widehat{Z} \} = 2 \widehat{m} $ & $ \{ \widehat{n}, \widehat{Z} \} = -2 \widehat{n} $ & & $ \{ \widehat{m}, \widehat{m} \} = 2 \widehat{m} $ & $ \{ \widehat{n}, \widehat{n}\} = 2 \widehat{n} $ \\
            \\
            $ \{ \widehat{\sigma}^{\dag}, \widehat{Y} \} = i \widehat{I} $ & $ \{ \widehat{\sigma}, \widehat{Y} \} = - i \widehat{I} $ & & $ \{ \widehat{Y}, \widehat{Y} \} = $ & & $2 \widehat{I} $ \\
            $ \{ \widehat{X}, \widehat{\sigma} \} = $ & $ \{ \widehat{\sigma}^{\dag}, \widehat{X} \} = $ & $\widehat{I}$ & $ \{ \widehat{X}, \widehat{X} \} = $ & & $2 \widehat{I} $ \\
            $ \{ \widehat{\sigma}, \widehat{\sigma}^{\dag} \} = $ & & $\widehat{I}$ & $ \{ \widehat{Z}, \widehat{Z} \} = $ & & $ 2 \widehat{I} $ \\
            $ \{ \widehat{\sigma}, \widehat{Z} \} = $ & $ \{ \widehat{Z}, \widehat{\sigma}^{\dag} \} = $ & $\widehat{0}$ & $ \{ \widehat{Z}, \widehat{X} \} = $ & $ \{ \widehat{Y}, \widehat{Z} \} = $ & $\widehat{0}$ \\
            $ \{ \widehat{\sigma}, \widehat{\sigma} \} = $ & $ \{ \widehat{\sigma}^{\dag}, \widehat{\sigma}^{\dag} \} = $ & $\widehat{0}$ & $ \{\widehat{X}, \widehat{Y}\} = $ & & $\widehat{0}$ \\
            \\
            $ \{ \widehat{n}, \widehat{m} \} = $ & & $\widehat{0}$ & & & \\
        \end{tabular}
    \end{center}
    \label{table_commutators_anticommutators}
\end{table}


\end{document}